\documentclass[a4paper,11pt]{article}

\pdfoutput=1  
\usepackage{jinstpub}
\usepackage{caption}
\usepackage{subcaption}
\usepackage{graphicx}
\usepackage{xcolor}
\definecolor{felix}{RGB}{0, 101, 189}
\definecolor{christian}{RGB}{41, 49, 123}
\definecolor{darkgreen}{RGB}{0, 150, 0}
\definecolor{darkblue}{RGB}{0, 0, 150}
\newcommand{\rev}[1]{{#1}}
\usepackage{tikz} % custom drawing
\usetikzlibrary{shapes,arrows,positioning,automata,backgrounds,calc,er,patterns,decorations.markings, shapes.misc}
\usepackage[compat=1.1.0]{tikz-feynman} % feynman diagrams
\usepackage{circuitikz} % draw circuits with latex :-)
\ctikzset{bipoles/length=.8cm}
\usepackage{booktabs}
\usepackage{multirow}
\usepackage{makecell}
\usepackage{framed}
\usepackage{physics} 
\usepackage{units} 
\usepackage[utf8]{inputenc}

\usepackage{lineno}
\title{STRAW (STRings for Absorption length in Water): pathfinder for a neutrino telescope in the deep Pacific Ocean}
\author[a]{M. Boehmer,}
\author[d]{J. Bosma,}
\author[d]{D. Brussow,}
\author[d]{L. Farmer,}
\author[a,1]{C. Fruck,}
\author[a]{R. Gernh\"auser,}
\author[a]{A. G\"artner,}
\author[b]{D. Grant,}
\author[a,f,1]{F. Henningsen,}
\author[a]{S. Hiller,}
\author[a]{M. Hoch,}
\author[a]{K. Holzapfel,}
\author[d]{R. Jenkyns,}
\author[a]{Na. Khera,}
\author[a]{Ni. Khera,}
\author[a]{K. Krings,}
\author[b]{C. Kopper,}
\author[d]{I. Kulin,}
\author[a]{K. Leism\"uller,}
\author[d]{J. Little,}
\author[d]{P. Macoun,}
\author[e]{J. Michel,}
\author[d]{M. Morley,}
\author[a]{L. Papp,}
\author[d]{B. Pirenne,}
\author[d]{C. Qiu,}
\author[a]{I. C. Rea,}
\author[a,1]{E. Resconi,\note{Corresponding author.}}
\author[d]{A. Round,}
\author[d]{A. Ruskey,}
\author[a]{C. Spannfellner,}
\author[c]{M. Traxler.}

\affiliation[a]{Physik-Department, Technische Universit\"at M\"unchen, D-85748 Garching, Germany}
\affiliation[b]{Dept. of Physics, University of Alberta, Edmonton, Alberta, Canada T6G 2E1}
\affiliation[c] {Helmholtzzentrum f\"ur Schwerionenforschung (GSI)
Planckstrasse 1, 64291 Darmstadt, Germany}
\affiliation[d]{Ocean Networks Canada, University of Victoria, Victoria, British Columbia, Canada}
\affiliation[e]{Institut f\"ur Kernphysik, Goethe Universit\"at, 60438 Frankfurt}
\affiliation[f]{Max-Planck-Insitut f\"ur Physik, D-80805 Munich, Germany}

\emailAdd{cfruck@ph.tum.de}
\emailAdd{felix.henningsen@tum.de}
\emailAdd{elisa.resconi@tum.de}

\abstract{\rev{The instrumentation for a} pathfinder mission \rev{towards} a possible large scale neutrino telescope named ``STRings for Absorption length in Water'' (STRAW) \rev{is presented in terms of design and performance}. In June 2018 STRAW \rev{was} deployed at the Cascadia Basin site operated by Ocean Networks Canada and has been collecting data since then. At a depth of about 2600 meters, the two STRAW 120 meters tall mooring lines are instrumented by three ``Precision Optical Calibration Modules'' (POCAM) and five Digital Optical Sensors (sDOM). \rev{The main objectives of STRAW are the measurement of light extinction in different wavelength bands and bioluminescence at Cascadia Basin.} We describe the instrumentation \rev{deployed in the Pacific Ocean and show some data from the first measurements.}}

\keywords{Neutrino detectors; Cherenkov detectors; \rev{detectors for astroparticle physics}}%Large detector systems for particle and astroparticle physics.}

\begin{document}
\maketitle
\flushbottom

\section{Introduction}
\label{sec:intro}
The Sun, core-collapse supernovae, and high-energy proton accelerators present in the cosmos are the most sophisticated cosmic laboratories nature has provided us for studying the universe over many orders of magnitude in energy, matter density and path length. While challenging to detect, neutrinos are an ideal probe for such extreme environments, with intervening matter, magnetic fields, and background radiation not affecting their travel from the sites of production to detection.  In particular, at the highest energies, above the TeV scale, the sky also begins to become opaque to photons. This makes neutrinos a particularly valuable astronomical messenger, providing a new window on the universe that has been recently opened to new phenomena at the highest energies thanks to the IceCube neutrino telescope based at the South Pole \cite{Aartsen:2013jdh, IceCube:2018dnn, IceCube:2018cha}. 

%To collect a statistically significant number of astrophysical neutrinos, huge volumes (cubic-kilometer scale) of transparent material must be instrumented with photon (and possibly acoustic/radio) sensors\,\cite{Markov:1961tyz}.  During the 1970s, the Deep Underwater Muon and Neutrino Detection (DUMAND) project\,\cite{Babson:1989yy} at 4800 m depth in the Pacific Ocean off the Big Island of Hawaii, initiated  a visionary exploration of the deep ocean for the first large volume neutrino telescope. The DUMAND project faced numerous challenges of deep water operation, and following 20 years of attempted deployment operation the project was cancelled. It readily recognized that  many of the lessons learned from the DUMAND project and its vision for high-energy neutrino astronomy has inspired a number of small- and medium-scale neutrino telescopes in deep water: NESTOR\,\cite{Rapidis:2009zz}, ANTARES\,\cite{Aguilar:2006rm}, and NEMO\,\cite{Adrian-Martinez:2015lef}, in the Mediterranean Sea, and the Baikal experiment in Siberia\,\cite{Belolaptikov:1997ry}. 
%The Baikal experiment and ANTARES have been successfully operated for various years  proving the technical feasibility of  permanent deep water neutrino telescope, though with a too small sensitive volume for detection of an astronomical signal up to now.

To collect significant numbers of astrophysical neutrinos, huge volumes (cubic-kilometers) of transparent material must be instrumented with photon (and possibly acoustic/radio) sensors\,\cite{Markov:1961tyz}.  During the 1970s, the Deep Underwater Muon and Neutrino Detection (DUMAND) project\,\cite{Babson:1989yy} was deployed at a depth of 4800~m in the Pacific Ocean off the Big Island of Hawaii. DUMAND initiated a visionary exploration of the deep ocean for the purpose of operating the first large volume neutrino telescope. Facing numerous challenges over 15 years of \rev{developments towards} deep water operation \rev{and after} a single deployment \rev{in 1993}, the project was ultimately cancelled.
It is readily recognized, however, that many of the lessons learned from the DUMAND project and its vision for high-energy neutrino astronomy inspired a number of small- and medium-scale neutrino telescopes in deep water, including:
%Aguilar:2006rm
NESTOR\,\cite{Rapidis:2009zz}, \rev{ANTARES\,\cite{ageron_antares:_2011}}, and NEMO\,\cite{Adrian-Martinez:2015lef}, in the Mediterranean Sea, and the Baikal experiment in Siberia\,\cite{Belolaptikov:1997ry}.
Though with a too small to be significantly sensitive to the detection of an astronomical signal, the Baikal and the ANTARES experiments have been successfully operating for the past decade, demonstrating the technical feasibility of a permanent deep-sea neutrino telescope. This success has not come without the challenges related to deep water operation, however, which has affected each of these projects in different ways. This has ultimately limited the ability to date to scale to the cubic-kilometer volumes needed to perform neutrino astronomy.
%The challenges related to deep water operation has affected each of these projects in different ways, limiting the ability to scale to the cubic-kilometer volumes needed to perform neutrino astronomy.

In the current era, the only cubic-kilometer neutrino telescope in operation \rev{worldwide} is the IceCube South Pole neutrino telescope\,\cite{Aartsen:2016nxy}. The detector is installed in the deep optically pristine Antarctic ice; a nearly ideal Cherenkov medium. With the observation of the first non-stellar neutrino source\, \cite{IceCube:2018dnn, IceCube:2018cha}, 
%the first Glashow resonance neutrino event [REF], and the first high-energy tau neutrinos [REF],
IceCube in its 10$^\mathrm{th}$ year of operation has launched a new discovery era in multi-messenger astronomy. To move from these first measurements to actually performing ``routine'' astronomical observations with neutrinos, it will be necessary to improve the current IceCube sensitivity to astrophysical phenomena by nearly two orders of magnitude.  This is a daunting challenge for any single facility to achieve, and the international community is investigating a global network of multi-cubic-kilometer neutrino telescopes to achieve this goal. %What does such a new landscape in multi-messenger neutrino astronomy look like?

In \cite{fa6bad185b2b4c48ad9f1d649717cd4a}, the IceCube collaboration presents a vision for a substantial expansion of the current IceCube detector, IceCube-Gen2, including  an instrumented array of an approximately 10\,km$^3$ volume of clear glacial ice at the South Pole to deliver a factor of 5 increases in the astrophysical neutrino sample with respect to the IceCube one. Given the complexity of the logistics and installation at the South Pole, the complete IceCube-Gen2 neutrino telescope is envisioned to be operative by 2032-2035. 

KM3NeT\footnote{https://www.km3net.org} is a European research infrastructure under construction in Italy and in France. The ultimate vision of KM3NeT is to have a detector volume of several cubic kilometers of clear sea water. While the mass production of the state-of-the-art photo-sensors and mooring lines has proceeded well, the installation of the \rev{deep-sea} infrastructure has continued to present \rev{several} technical challenges \rev{that have}, thus far, \rev{considerably delayed} the deployment and operation of the detector arrays.

%http://baikalweb.jinr.ru
The Baikal Gigaton Volume Detector (Baikal-GVD)\footnote{http://inspirehep.net/record/1339487} is optimized to detect neutrinos at energies from a few TeV up to 100 PeV. Once completed, it will instrument a bit more than 1\,km$^3$ of the deep water of Lake Baikal.

New in this constellation of sites is the Ocean Networks Canada (ONC). ONC operates undersea sensor networks around Canada. It is an initiative of the University of Victoria, in British Columbia. The network infrastructure supports interdisciplinary science through a wide variety of instruments connected by undersea cables that provide ample power and gigabits per second of communication capability. The main observatories operated by ONC are called NEPTUNE and VENUS, and are located in the northeast Pacific Ocean and the Salish Sea, respectively. Of particular note, the ONC deep-sea infrastructure at Cascadia Basin dark site (47$^\circ$46'\,N, 127$^\circ$46'\,W), at a depth of $\sim$2600\,m b.s.l., provides a number of the ideal prerequisites for a possible large scale neutrino telescope. 
To this end, since late 2017 a pathfinder mission has emerged named ``STRings for Absorption length in Water'' (STRAW) with the aim to build and deploy photodetectors and calibration instrumentation. In June 2018 STRAW \rev{was} deployed at the Cascadia Basin site and has been collecting data since then.

The goal of STRAW is a systematic, step-by-step investigation of the {\it in-situ} optical transparency and ambient background light of the site.  Operation of the two 120\,m tall mooring lines is on-going and anticipated through 2020 to provide long-term monitoring of the site's optical properties.  Included in the summer 2018 deployment was an {\it in-situ} spectrophotometer (WetLabs AC-9) that provided a complementary measure of the optical properties as a function of wavelength and depth.  The preliminary results from the STRAW studies, including the ease of deployment and operation, indicate significant potential for the site to host a future large-volume neutrino telescope. 

\rev{This paper starts with a description of the overall concept of STRAW, followed up by a detailed description of the two main functional modules: Precision Optical Calibration Module (POCAM) and STRAW Digital Optical Module (sDOM). After describing the mechanical design of the strings and the seafloor infrastructure, we report on tests that were performed on the modules, before we report on the deployment itself. We conclude this paper with some first measurements recorded with STRAW.}

\section{STRAW Concept and Design}
\label{sec:design}
%Note for us: Extinction (or attenuation) is the sum of scattering and absorption, so it represents total effect of medium on radiation passing the medium. We measure extinction (or attenuation) but we aim to quote scattering and absorption length.
With the vision of a new large scale neutrino facility in the northern Pacific\,\cite{Vallee:2016xde}, we have built and deployed the STRAW pathfinder (see Fig.\,1). %Fig.\,\ref{fig:STRAW-scheme} WRONG REFERENCE???). 
%The availability of the deep-see infrastructure provided by ONC at Cascadia Basin (47$^\circ$46'~N, 127$^\circ$46'~W), together with the convenient depth of $\sim$ 2600~m.b.s.l.,  provides ideal perquisites for a possible large scale neutrino telescope.
The main tasks of STRAW are to (1) measure the absorption and scattering length of the deep-sea water at Cascadia Basin in the wavelength range between 350\,nm and 600\,nm, and (2) characterize the overall ambient background light produced by the bioluminescence of deep-sea living organisms and the $^{40}$K dissolved in the salty water. 
STRAW  is  designed following the principle of neutrino telescopes which includes the use of PMT light sensors, LED calibration light sources, and specific choices about the readout electronics and deployment strategy.
\begin{figure}[ht]
\centering
\includegraphics[height=0.9\textheight]{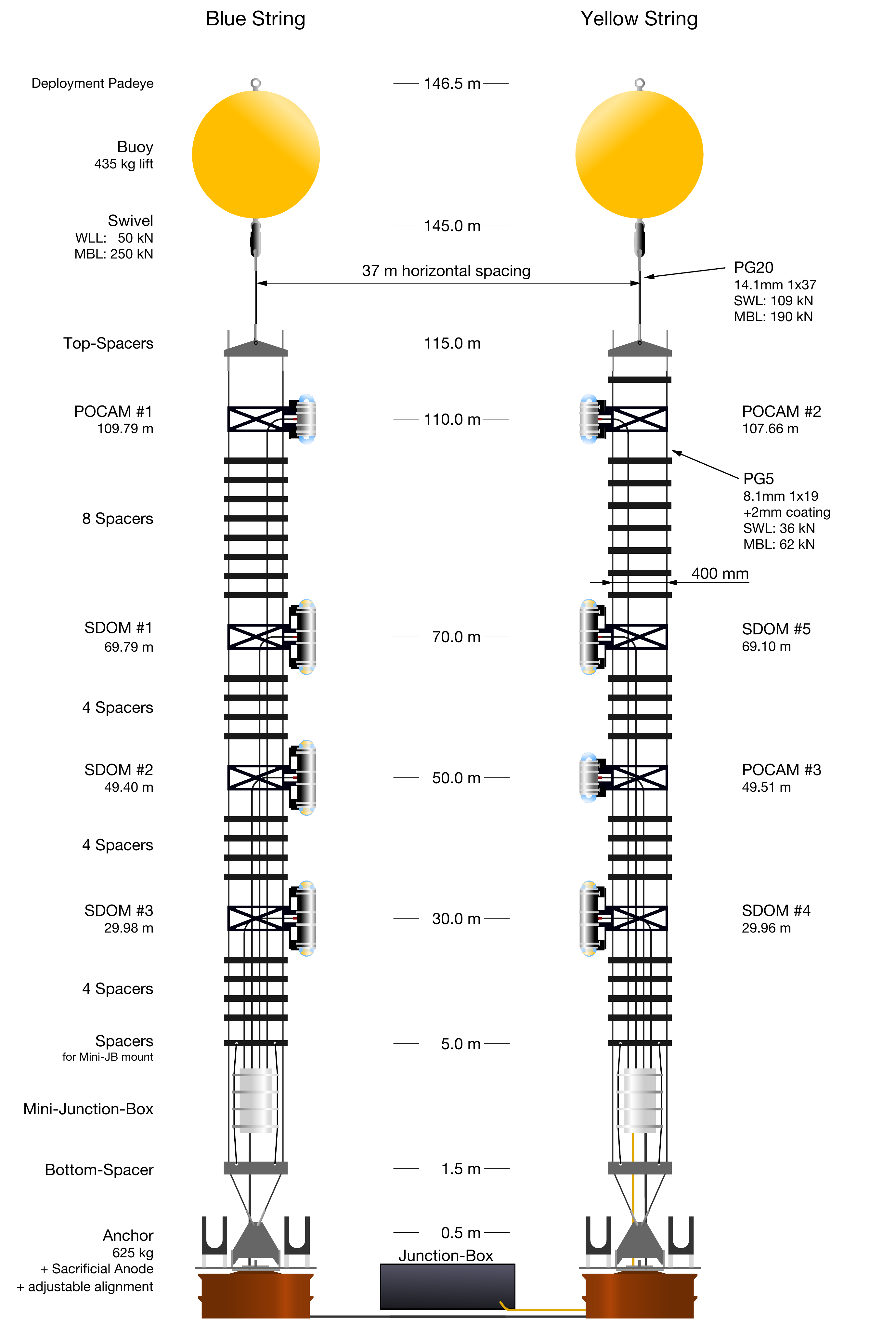}
\label{fig:STRAW-scheme}
\caption{Detailed technical sketch of the two STRAW mooring lines showing the exact (measured) geometry of all modules.}
\end{figure}
%
%DESIGN
STRAW consists of two vertical mooring lines instrumented with {\it light emitter modules} and {\it light sensor modules} mounted at different heights above the sea floor. The light emitter module is based on the design of 
the {\it Precision Optical Calibration Module} (POCAM)\,\cite{Jurkovic:2016kxn}, which provides an 
isotropic and short pulsed flash of light (see Sec.\,\ref{sec:POCAM}).
The light sensor module named the \textit{STRAW Digital Optical Module} (sDOM) is using two 3$^{\prime\prime}$ PMTs encapsulated in a titanium cylinder similar to the one of the POCAM. The entire PMTs' readout and data acquisition system is also included in the module (see Sec.\,\ref{sec:sDOM}).

The measurement of the attenuation length $L_T$ in water will be realized by emitting a flash of photons of known intensity $N_0$ and wavelength from one of the POCAM units and detecting it at a known distance $r$ in one of the sDOMs with effective collection area $A_{\mathrm{det.}}$, measuring an intensity $N(r)$. 
\begin{equation}
	N  \, (r) = \frac{ N_0 }{4\pi r^2} \; \exp(- \frac{r}{L_T}) \; A_{\mathrm{det.}}.
	\label{eq:light_extinction}
\end{equation}
From theoretical predictions \citep{mobley_light_1994} and measurements in pure water \citep{smith_optical_1981} as well as sea water \citep{capone_measurements_2002} it is known that the absorption length is highest for the wavelength band around 460\,nm and decreases quickly in the UV as well as towards 600\,nm. Fortunately, this is also the most sensitive wavelength band for the detection of Cherenkov light, especially when using PMTs. Expected maximum values for the absorption length are around 50\,m; therefore a geometry has been chosen that covers distances from 20\,m to 90\,m. The geometry also incorporates a fair amount of symmetry to allow for a study of systematic uncertainties on a module to module basis. Other limitations on the geometry of the 2-string array were imposed by technical constraints like the maximum safe cable length for Ethernet and RS485 connections without the danger of data loss over twisted-pair Copper wires, 70\,m, and 130\,m respectively. Another limitation was introduced by \textit{remotely operated vehicle} (ROV) operations during the deployment, limiting the minimum safe distance for deployment to a distance of around 40\,m.

\section{The Precision Optical Calibration Module (POCAM)}
\label{sec:POCAM}
The POCAM was initially designed to act as a light emitting calibration device for IceCube to improve the understanding of the optical ice properties\,\cite{Jurkovic:2016kxn}. Using LEDs of different wavelengths, it can emit isotropic light flashes with known intensity due to the implementation of {\it in-situ} self-calibration using integrated photosensors. In STRAW, the POCAMs are used for emitting intense ($\mathcal{O}(10^9)$ photons), \rev{adjustable (via the supply voltage)}, isotropic, nanosecond light flashes that are observed by the surrounding sDOM units. One major challenge when using LEDs was to achieve a high level of intensity and isotropy with these flashes while keeping the nanosecond timing. The former is necessary to illuminate a large detector volume homogeneously, and the timing allows differentiation between direct and scattered light. \rev{The adjustable light intensity is required in order to be able to be able to detect single photons over a large range of distances and for different possible attenuation lengths}.

As the development of the POCAM was part of a Master Thesis, many of the following figures were adapted from \citep{Henningsen:2019jor}. A lot of additional and more detailed information on the POCAM characteristics can also be found there.

\subsection{Housing}
\begin{figure}[h!]
    \centering
	\includegraphics[angle=0,origin=\textwidth,width=0.45\textwidth]{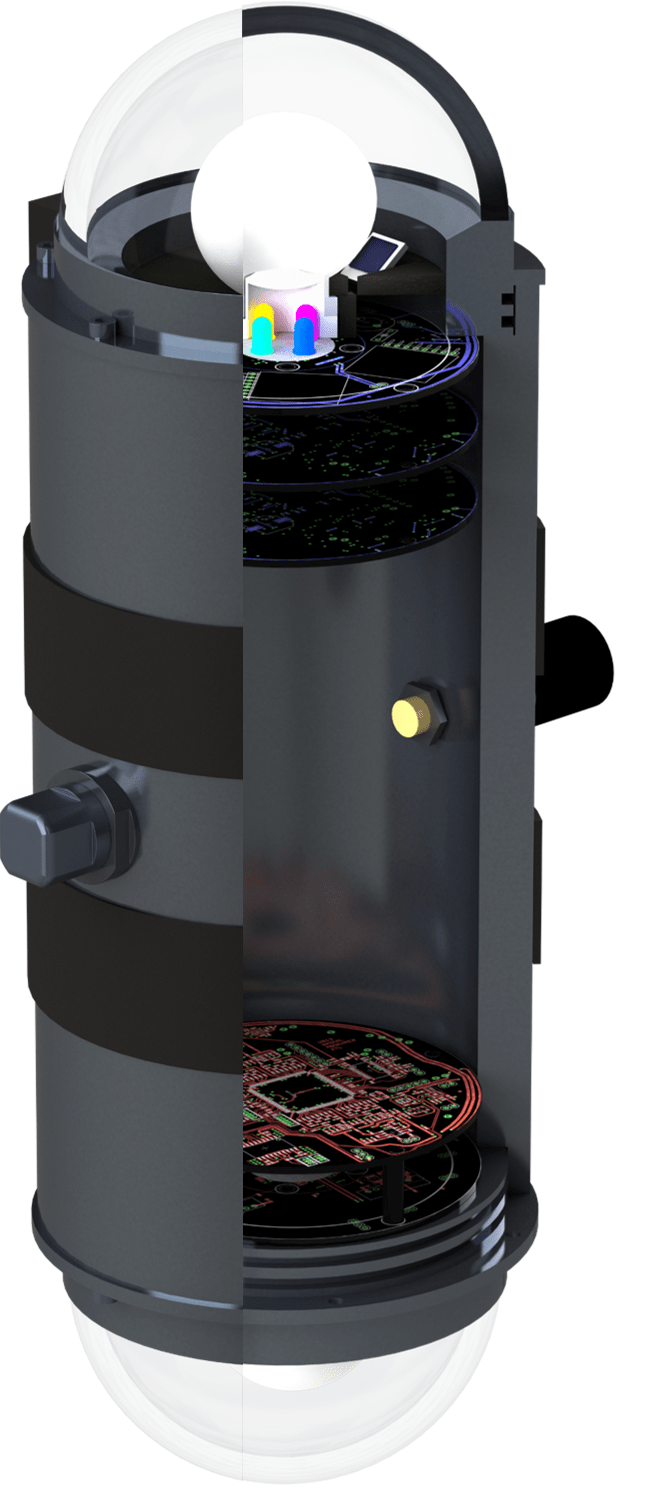}
	\caption{The POCAM housing. Titanium cylinder with BK-7 glass hemispheres on either end allowing complementary, isotropic light emission.}
	\label{fig:pocam-housing}
\end{figure}

Since the neutrino telescopes are usually deployed deep below the surface in water or ice, the housing of the instruments has to withstand high pressures of several hundred atmospheres. For practical reasons and in order to allow for good optical transmission this is typically achieved by using glass pressure spheres in which the photosensors and electronics are encapsulated and protected. Initially, this was also the idea for the POCAM, but simulations of the light diffuser indicated that this component needs to be small in order to avoid stretching of the light pulses\,\cite{veenkamp:thesis:2016}. Also in order to create the necessary space for the electronics, the cylindrical housing design shown in Fig.\,\ref{fig:pocam-housing} was used.
\par
In order to achieve a high-pressure resistance and at the same time avoid corrosion, the cylindrical part of the housing was made from titanium. For the glass hemispheres, an optically enhanced borosilicate glass (N-BK7) has been used \rev{which provides a transmissivity of >95\% in the range between 350\,nm and 600\,nm\,\cite{nbk7}}, at \rev{significantly} reduced costs compared to quartz. Borosilicate glasses are also less prone to temperature shock damage, as they come with low thermal expansion, which is beneficial for a potential application in IceCube. The glass itself is attached to titanium flanges using deep-sea-grade epoxy resin.
A vacuum port allows for degassing and nitrogen-flushing of the instrument, which removes air and subsequently humidity. This greatly reduces the risk of condensation that could cause electrical problems. The housing itself is kept at approximately 0.3$\,$bar also to ensure sealing in low-pressure water environments like test tanks. Lastly, electrical supply and control are enabled by a connector located in the center of the cylinder. According to the manufacturer, this design will withstand an \rev{external} pressure of 1500$\,$bar and temperatures down to -40$\,^\circ$C. As will be discussed later, the housing was subjected to and passed a number of environmental tests.
\subsection{Flasher Circuit}
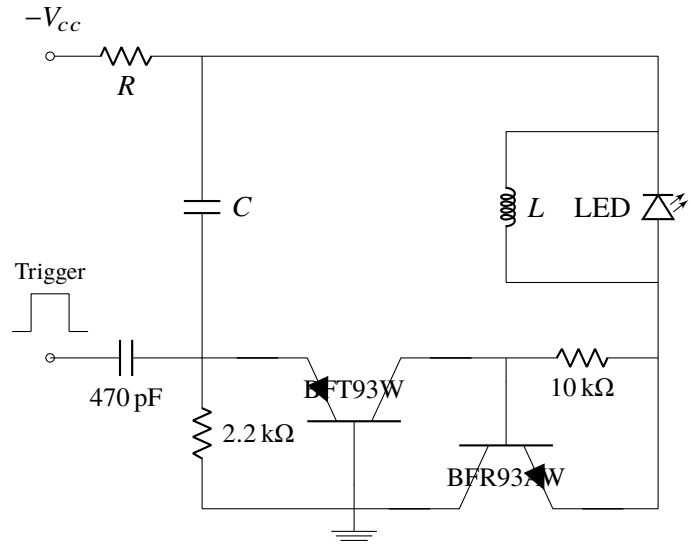
\begin{figure}[ht!]
	\centering
	\begin{circuitikz}[american voltages]
		\draw
  			(0,0) node[ocirc] {};
		% trigger input part
		\draw		
  			(0,0) 
  			to [capacitor, l_=\small $470\,$pF] (2,0)
  			to [capacitor, l_= $C$] (2,4)
  			to [resistor, a^= $R$] (0, 4);
  		\draw
  			(-0.5,0.35)
  			to [short] (-0.25, 0.35)
  			to [short] (-0.25,0.85)
  			to [short] (0.25,0.85)
  			to [short] (0.25,0.35)
  			to [short] (0.5,0.35);
  		\draw
  			(0, 1.125) node[] {\footnotesize Trigger};
  		% VCC
  		\draw
  			(0,4) node[ocirc] {}
  			(0,4) node[above=2mm] {~$-V_{cc}$};
  		% general circuitry
  		\draw
  			(2,4) 
  			to [short] (8,4)
  			to [led, invert, l_=LED] (8, 0)
  			to [short] (8,-2)
  			to [short] (7, -2);
  		\draw
  			(8,3)
  			to [short] (6,3)
  			to [cute inductor, a^=$L$] (6,1)
  			to [short] (8,1);
  		\draw
  			(2,0)
  			to [short] (3,0);
  		% pnp
  		\draw
  			(4,0) node[pnp, scale=2, rotate=90](pnp) {}
  			(pnp.base) node[anchor=south]{}
  			(pnp.collector) node[anchor=west] {}
  			(pnp.emitter) node[anchor=east] {}
  			(pnp.base) node[right=5mm, above=10mm]{\small BFT93W};
  		\draw
  			(3,0)
  			to [short] (pnp.emitter);
  		\draw
  			(5,0)
  			to [short] (pnp.collector);
  		\draw
  			(4,-2)
  			to [short] (pnp.base);
  		% npn
  		\draw
  			(6,-2) node[npn, scale=2, rotate=270, yscale=-1](npn) {}
  			(npn.base) node[anchor=east]{}
  			(npn.collector) node[anchor=south] {}
  			(npn.emitter) node[anchor=north] {}
  			(npn.base) node[right=5mm, below=10mm]{\small BFR93AW};
 		\draw
  			(5,-2)
  			to [short] (npn.collector);
  		\draw
  			(6,0)
  			to [short] (npn.base);
  		\draw
  			(7,-2)
  			to [short] (npn.emitter);
  		\draw
  			(5,0)
  			to [short] (6,0)
  			to [resistor, l_=\small $10\,$k$\Omega$] (8,0);
  		% trigger resistor
  		\draw
  			(5,-2)
  			to [short]  (2,-2)
  			to [resistor, l_=\small $2.2\,$k$\Omega$] (2,0);
  		% ground
  		\draw
  			(4,-2) node[ground, scale=1.5] {};
	\end{circuitikz}
	\caption[Kapustinsky flasher schematic.]{Kapustinsky flasher schematic for the POCAM in STRAW. The circuit is operated on a negative bias voltage and produces a pulse when triggered by a square pulse signal. The light pulse is mainly shaped by the capacitor $C$ and the inductance $L$, as well as the LED itself.}
	\label{fig:kapustinsky-schematic}
\end{figure}

The LED flasher follows the design reported in\,\cite{Lubsandorzhiev:2004zh}, which originates from the one by J.S. Kapustinsky in 1985\,\cite{KAPUSTINSKY1985612}. For the POCAM in STRAW, the parameters of the components have been adjusted in order to achieve a pulse \rev{FWHM} of $<10\,$ns and maximum light intensity. The final design of the circuit is shown in Fig.\,\ref{fig:kapustinsky-schematic}.
\par
The flasher circuit is operated with adjustable, negative bias voltage and produces a light pulse upon being triggered with a square pulse signal. In order to create the light flash, the capacitor gets loaded at an adjustable voltage. The trigger signal opens with its rising edge the two transistors of the circuit, which stay open until the capacitor has completely discharged. The capacitor then discharges through the LED creating the light pulse of a certain duration. The inductance parallel to the LED cuts subsequently the pulse after a few nanoseconds and creating an opposite current in the LED~\cite{Lubsandorzhiev:2004zh}.

\begin{figure}[h!]
	\includegraphics[width=\textwidth]{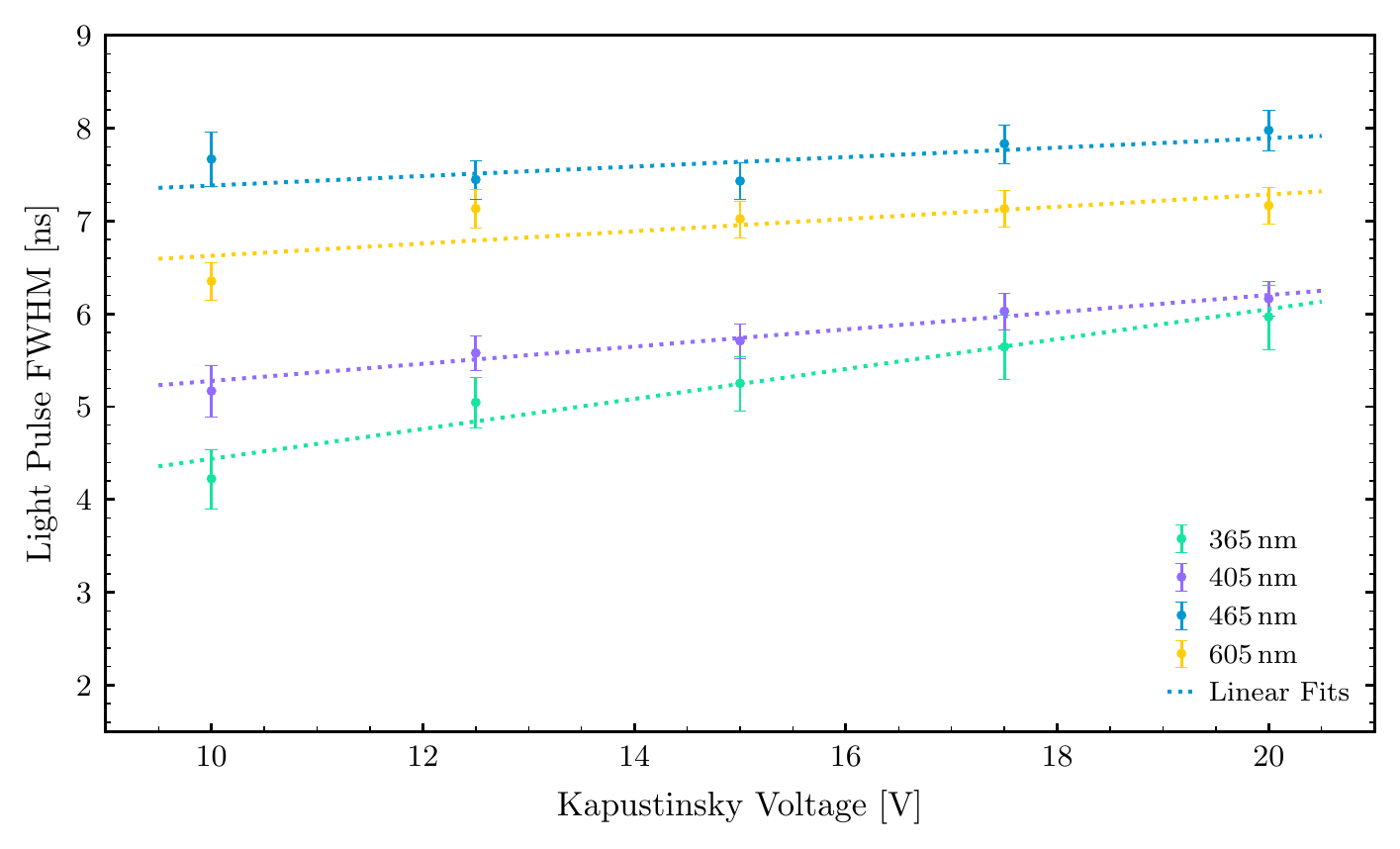}
	\caption{FWHM of the Kapustinski flasher pulse versus bias voltage.}
	\label{fig:fwhm}
\end{figure}

\begin{figure}[h!]
	\includegraphics[width=\textwidth]{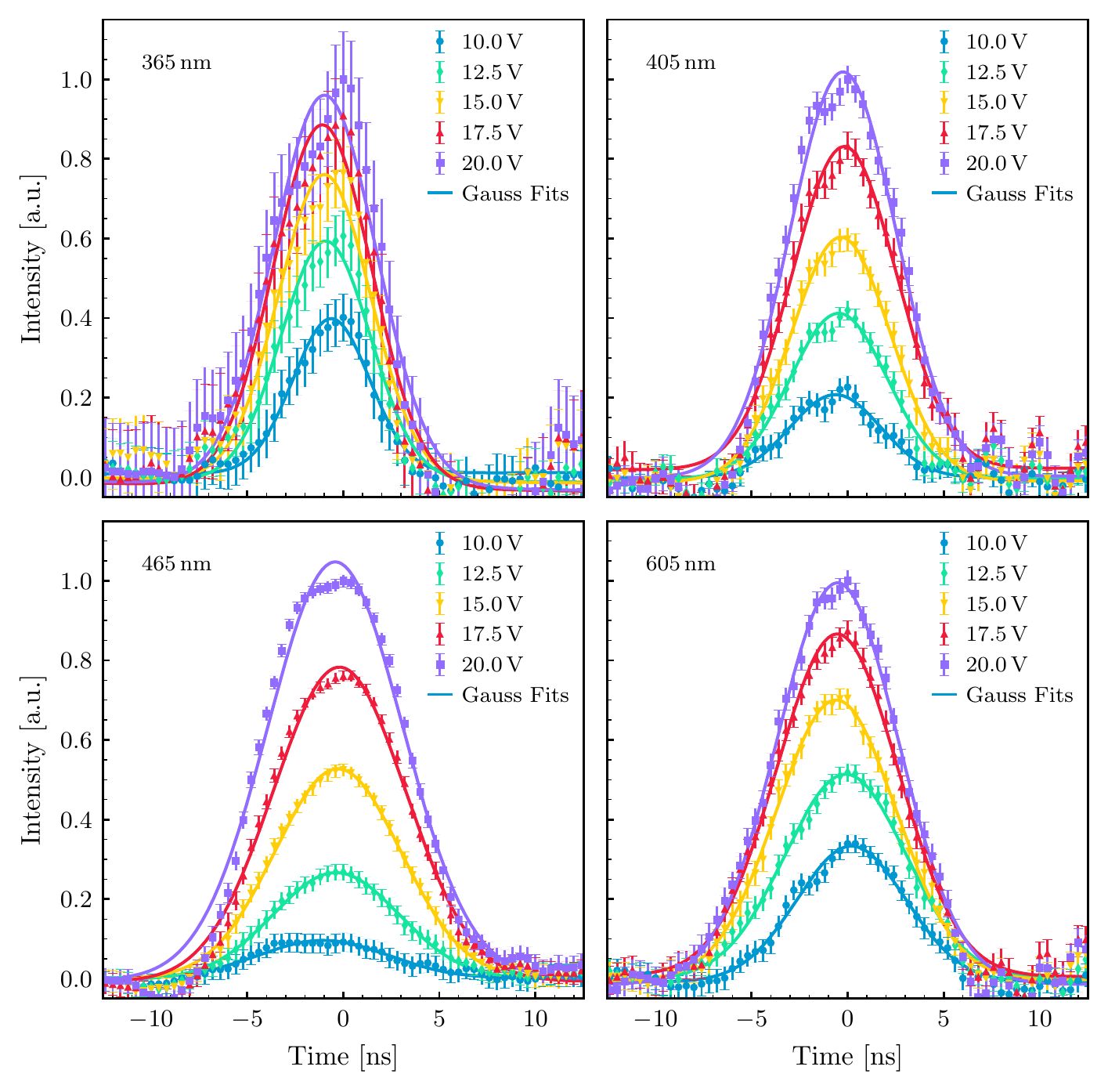}
	\caption{POCAM Pulse shapes for the selected Kapustinsky LEDs of indicated wavelengths in the top left corners. Shown are measurements made with a sub-nanosecond photodiode at different flasher voltages.}
	\label{fig:pulseshape}
\end{figure}

The exact shape and photon output of this pulse depend on the specific parameters of the capacitor $C$, the inductance $L$, the LED itself and the bias voltage. This means, larger inductance and capacitors will result in longer, brighter pulses, which was then used to optimize the flasher design of the POCAM in STRAW. During the development phase, it was found that using multiple LEDs in parallel increases the pulse intensity without impact on the timing. The final LED configuration is listed in Tab.\,\ref{tab:led-selection}. Extracted full widths at half maximum (FWHM) of the flashes and measured pulse-shapes are shown in Fig.\,\ref{fig:fwhm} and Fig.\,\ref{fig:pulseshape} respectively.
%\,Fig.\,\ref{fig:pulseshape}. 
In addition to the Kapustinsky flasher also two LEDs driven by the FPGA directly are included, which however only emit significant intensities at pulse widths of $40\,$ns and more. As such, they only act as backup flashers \rev{and pose the fallback solution in case the optical properties are much worse than expected as they can emit more light than the Kapustinsky flasher, but over a much longer time}. The final LED configuration is also shown in Fig.\,\ref{fig:LED-layout}.
%.\,\cite{Henningsen:2019jor}
\par
\begin{table}[ht!]
	\centering
	\begin{tabular}{l c c c c c }
		\toprule 
		\bf LED & \bf Color & \thead{\bf Spectral Range\\ \bf [nm]} & \thead{\bf Relative\\ \bf Intensity} & \bf Kapustinsky & \bf FPGA\\
		\midrule
		XSL-365-5E~\cite{365nm} & UV &  $365 \pm 8$ & $0.13$ & 1x & -- \\
		XRL-400-5E~\cite{400nm} & Violet & $405 \pm 8$ & $1.00$ & 1x  & --\\
		NSPB300B~\cite{470nm} & Blue & $465 \pm 7$ & $0.62$ & 2x & 2x\\
		WP710A10LZGCK~\cite{525nm}  & Green & $525 \pm 17$ & $0.23$ & -- & 2x\\
		CSL0701DT5~\cite{605nm} & Orange & $605 \pm 6$ & 0.65 & 2x & --\\
		\bottomrule
	\end{tabular} 
	\caption[Final selection of POCAM LEDs for STRAW.]{Final selection of the six POCAM LEDs for STRAW. Given are the LED part names, their given FWHM spectral ranges, their calibrated, relative intensities at $\unit[20]{V}$ Kapustinsky voltage and their multiplicities on the Kapustinsky and FPGA flashers, respectively.}
	\label{tab:led-selection}
\end{table}
\par
\subsection{Integrating Sphere}
\begin{figure}[h!]
    \centering
    % ptfe sphere
    \begin{subfigure}[b]{.49\textwidth}
    \centering
	\vspace{-12pt}
	\begin{tikzpicture}
   			\node[anchor=south west,inner sep=0] (image) at (0,0) {\includegraphics[width=0.6\textwidth]{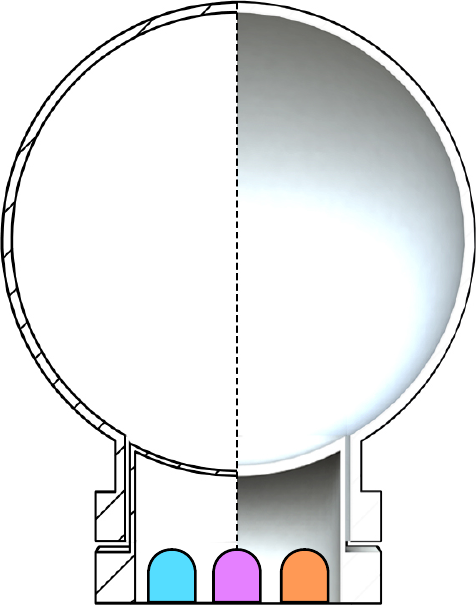}};
    		\begin{scope}[x={(image.south east)},y={(image.north west)}]
        		%\draw[help lines,xstep=.2,ystep=.1] (0,0) grid (1,1);
        		%\foreach \x in {0,1,...,9} { \node [anchor=north] at (\x/10,0) {0.\x}; }
        		%\foreach \y in {0,1,...,9} { \node [anchor=east] at (0,\y/10) {0.\y}; }
        		\draw (0.025, 0.1) -- (0.125, 0.1);
        		\draw (0.025, 0.09) -- (0.025, 0.11);
        		\draw (0.125, 0.09) -- (0.125, 0.11);
        		\node at (0.05, 0.05) {\footnotesize 5$\,$mm};
    		\end{scope}
		\end{tikzpicture}
	\caption{POCAM integrating sphere geometry in cut-view. Visible is the two-part sphere geometry with the plug and the sphere exterior. The neck hosts the LED on the bottom, flashing upwards.}
	\label{fig:ptfe-sphere}
    \end{subfigure}\hfill
    % flasher led layout
	\begin{subfigure}[b]{.49\textwidth}
	\centering
    \includegraphics[width=0.6\textwidth]{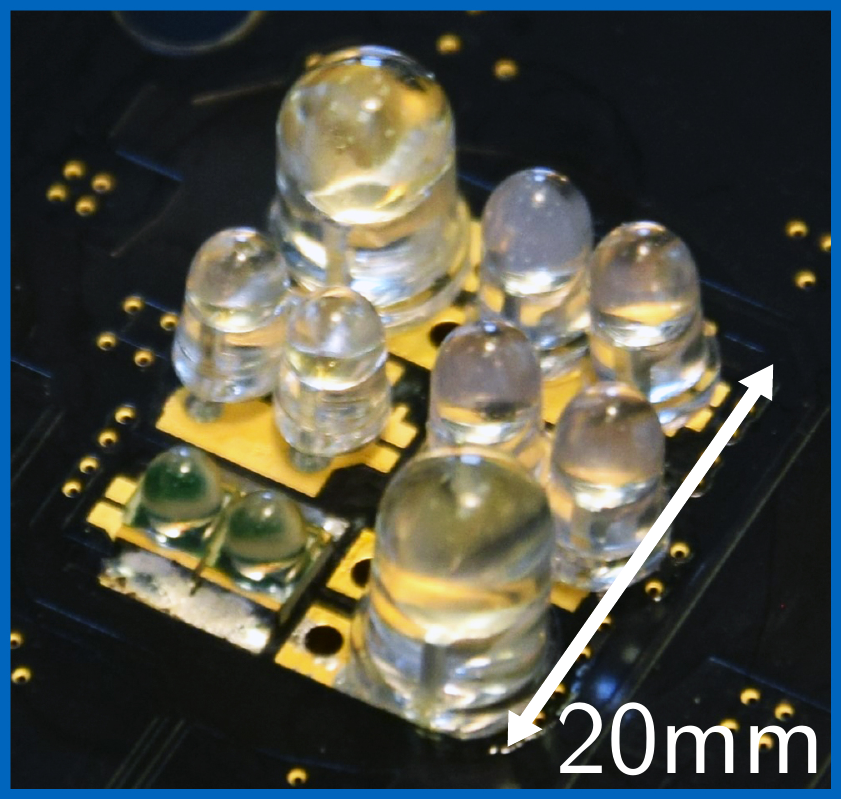}
	\caption{POCAM LED layout as used inside the integrating sphere plug. The corner LEDs are Kapustinsky-driven, the central LEDs are backup flashers driven by the FPGA directly.}
	\label{fig:LED-layout}
    \end{subfigure}
    \caption{POCAM integrating sphere (a) and LED layout (b).}
    %\label{fig:intsphere-flasher}
\end{figure}

LEDs typically emit light in a cone with an opening angle that varies from model to model. In the POCAM the isotropic light emission is realized by a custom made integrating sphere made by \textit{polytetrafluoroethylene} (PTFE)~\cite{ptfe-optics}.
The POCAM integrating sphere consists of two mechanical parts, as shown in Fig.\,\ref{fig:ptfe-sphere}. The emitting part is a sphere with a circular recess on the bottom and a neck. The second piece is a plug that completes the sphere interior for light integration and provides a pre-diffusing PTFE layer. With a diameter of only $50\,$mm, the light integration does not significantly alter the pulse shape as has been shown in previous studies~\cite{Andi:BA}.
\par
\begin{figure}[h!]
	\includegraphics[width=\textwidth]{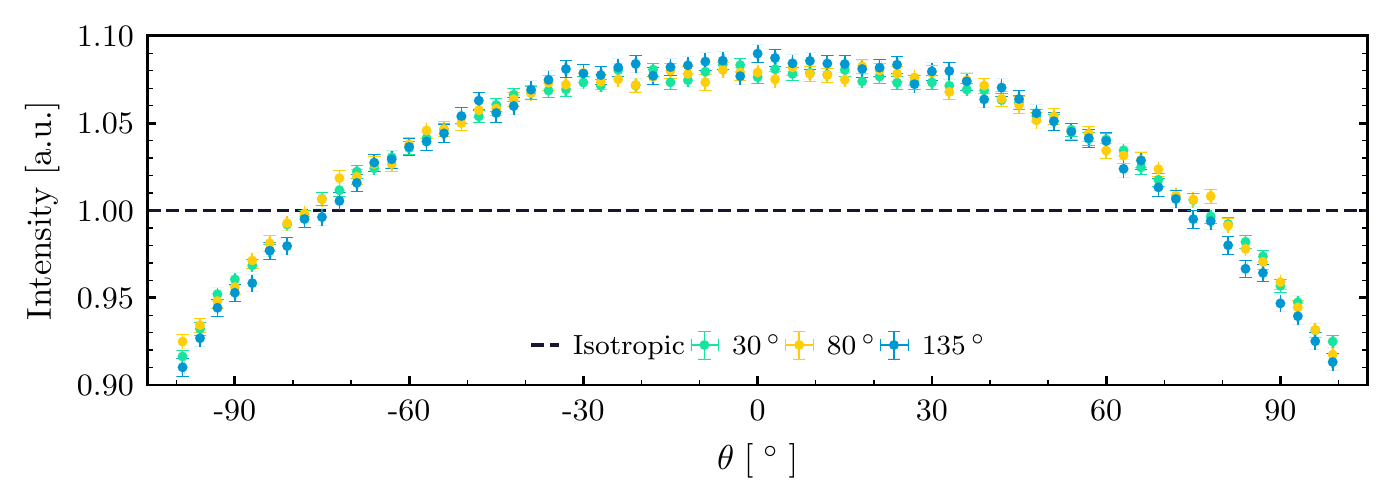}
	\caption{Isotropy scan of a POCAM integrating sphere. The polar angle is measured with respect to the axis of LED emission with $\theta=0$ being the pole of the sphere. The plot shows measurements of different LED opening angles and shows the effect of the diffusion resulting in equivalent emission profiles. The error bars arise from azimuthal deviations.}
	\label{fig:isotropy}
\end{figure}
\begin{figure}[h!]
    \hspace{-2mm}
	\includegraphics[width=\textwidth]{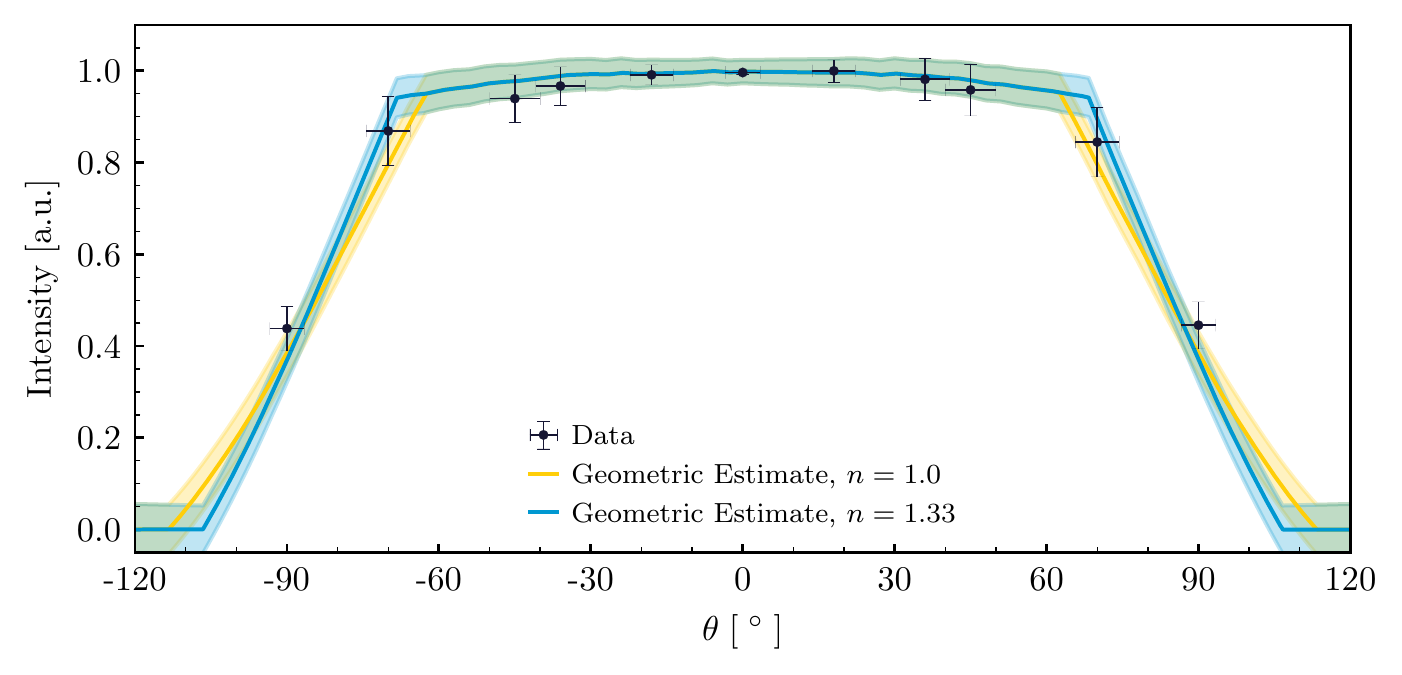}
	\caption{Geometric estimate of a POCAM hemisphere emission versus polar angle. Shown are measurements and geometric estimates on a hemisphere emission profile with air or water as external medium, respectively. Additionally, a preliminary measurement of the angular intensity in air is shown which is matched fairly well by the geometric estimate.}
	\label{fig:pocam-isotropy}
\end{figure}

The isotropy achieved with this geometry has been investigated using an automated dual-axis rotation measurement setup. The achieved emission profile is shown in~Fig.\,\ref{fig:isotropy} and shows that the integrating sphere successfully homogenizes flashes of different LED opening angles. Using this scan, a polar isotropy of $5.4\,\%$ and an azimuthal uniformity of $4.7\,\%$ was determined. Both prove feasible for a STRAW deployment. However, it also shows that the initial direction of the LED flash is not yet completely lost and the emission isotropized as wished. This is subject of further optimization studies that are currently ongoing.
The POCAM emission including its housing geometry and refractive effects is reported in~Fig.\,\ref{fig:pocam-isotropy}. 
\par
\par

\subsection{Electronics}
The functionality of the POCAM is implemented on three different boards: 
\begin{itemize}
    \item the \textit{power supply board} converts the VDC input voltage to local 12\,VDC; it provides a local 10\,MHz clock to both hemispheres;
    \item the \textit{digital board} which hosts a microcontroller, an FPGA, a 10-bit/10\,MHz ADC and storage;
    \item the \textit{analog board} which hosts the LED flashers, the integrated photosensors and the amplifier chain necessary for their readout. 
\end{itemize}
The POCAM is controlled by a half-duplex RS485 serial line. Commands are handled by the microcontroller, controlling the FPGA for real-time tasks. Acquired data is stored onboard and transferred after flashing to the surface.

Special care has been taken to keep the system accessible from the surface: both microcontroller and FPGA feature a golden image to allow recovery in case of failure; microcontroller's bootloader allows reflashing it by RS485 lines in addition.

\begin{figure}[h!]
	\includegraphics[width=\textwidth]{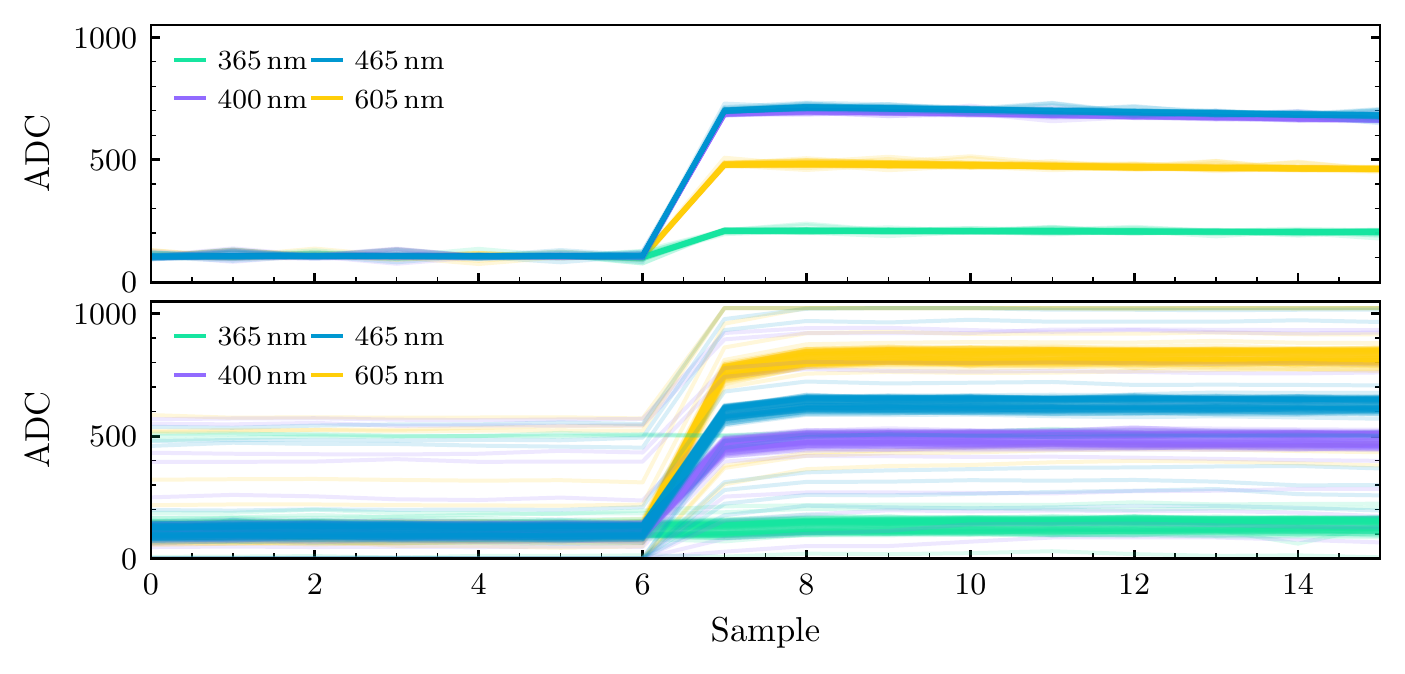}
	\caption{The readout of the POCAM light pulse is monitored {\it in-situ} by two photosensors a SiPM (top) and a photodiode (bottom). The abscissa shows the samples which are 100\,ns steps of the internal clock. The ordinate shows the ADC readout value in bits.}
	\label{fig:pocam-raw}
\end{figure}

Two integrated photosensors,  Silicon-Photomultiplier (SiPM) and a photodiode,  provide the {\it in-situ} monitoring of the light emitted.   The combination of these two sensors allows the monitoring of a high light dynamic range: as the SiPM is sensitive to faint amounts of light, the photodiode is robust to very high light intensity.  The outputs of both sensors are guided through an amplifier chain and fed into a charge-amplifier which integrates the charge and produces a proportional output voltage with a decay time several tens of microseconds. These slow signals are then sampled by the 10-bit ADC and stored locally. Exemplary outputs of these measurements are shown in~Fig.\,\ref{fig:pocam-raw}.
\par
\subsection{Internal Structure and Shadow Mount}
The circuit boards are mounted on circular boards, which are stacked \rev{using plastic support pillars for PCBs (Printed Circuit Board).} The structure is attached to the titanium flange of the housing to provide enough mechanical stability. 
Additionally, a \textit{shadow mounting} is placed below the integrating sphere and above the analog board to remove possible reflections of its surface. It is coated with an highly-absorbing spectrometer paint. The shadow mounting has drill holes which define constant solid angles for the photosensors located below. To verify the mechanical stability of the internal structure of the POCAM, a standardized shock- and vibrational testing procedure \rev{was applied (see Sec.\,\ref{sec:testing})}.

\subsection{POCAM Absolute Calibration}
\begin{figure}[h!]
	\includegraphics[width=\textwidth]{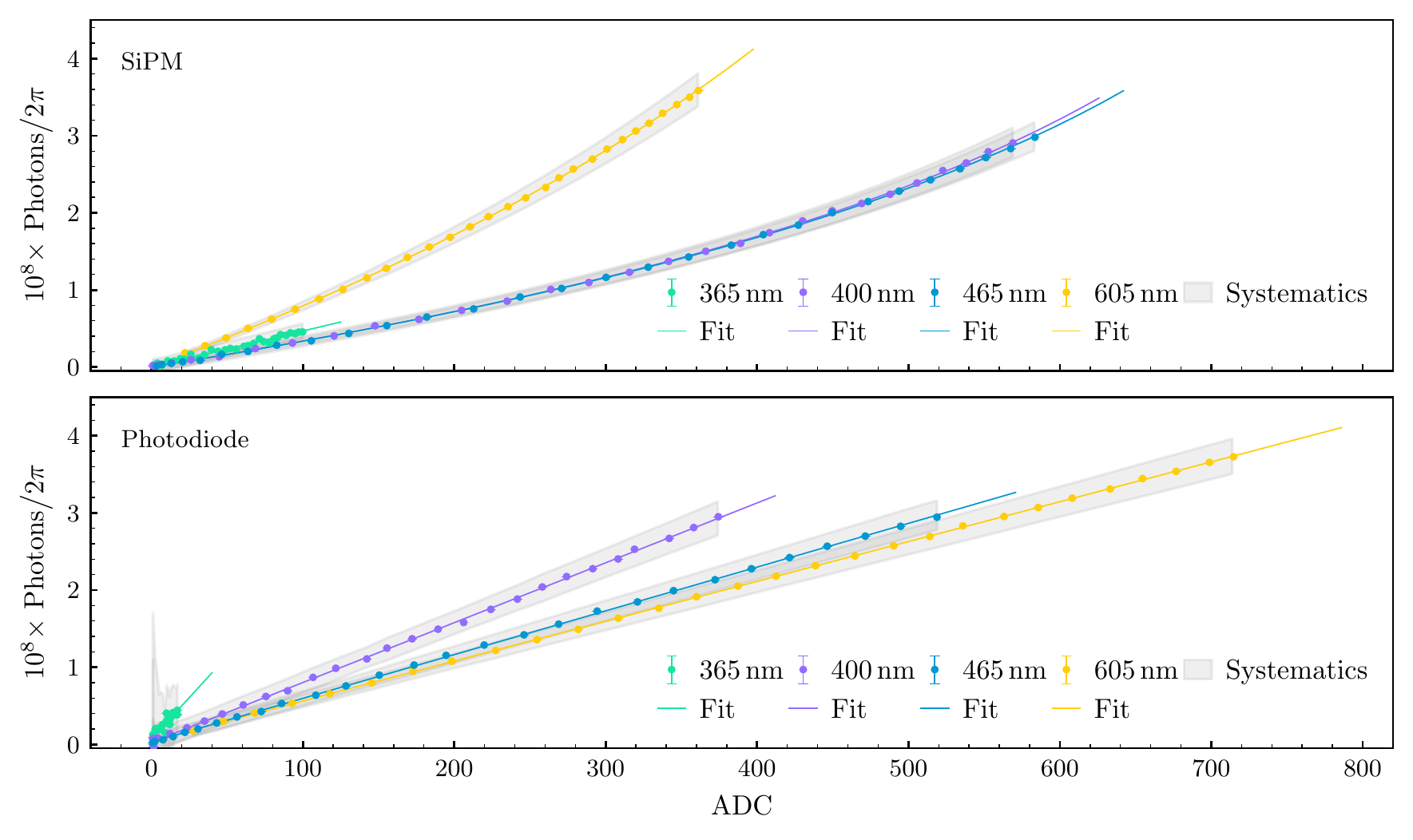}
	\caption{POCAM light output calibration. Shown is the total number of photons extrapolated from the reference measurement and the emission profile versus the internal ADC readout which corresponds to different flasher circuit voltages including fits. Error bars are statistical, the grey band shows systematic uncertainties.}
	\label{fig:pocam-calibration}
\end{figure}

\begin{figure}[h!]
	\includegraphics[width=\textwidth]{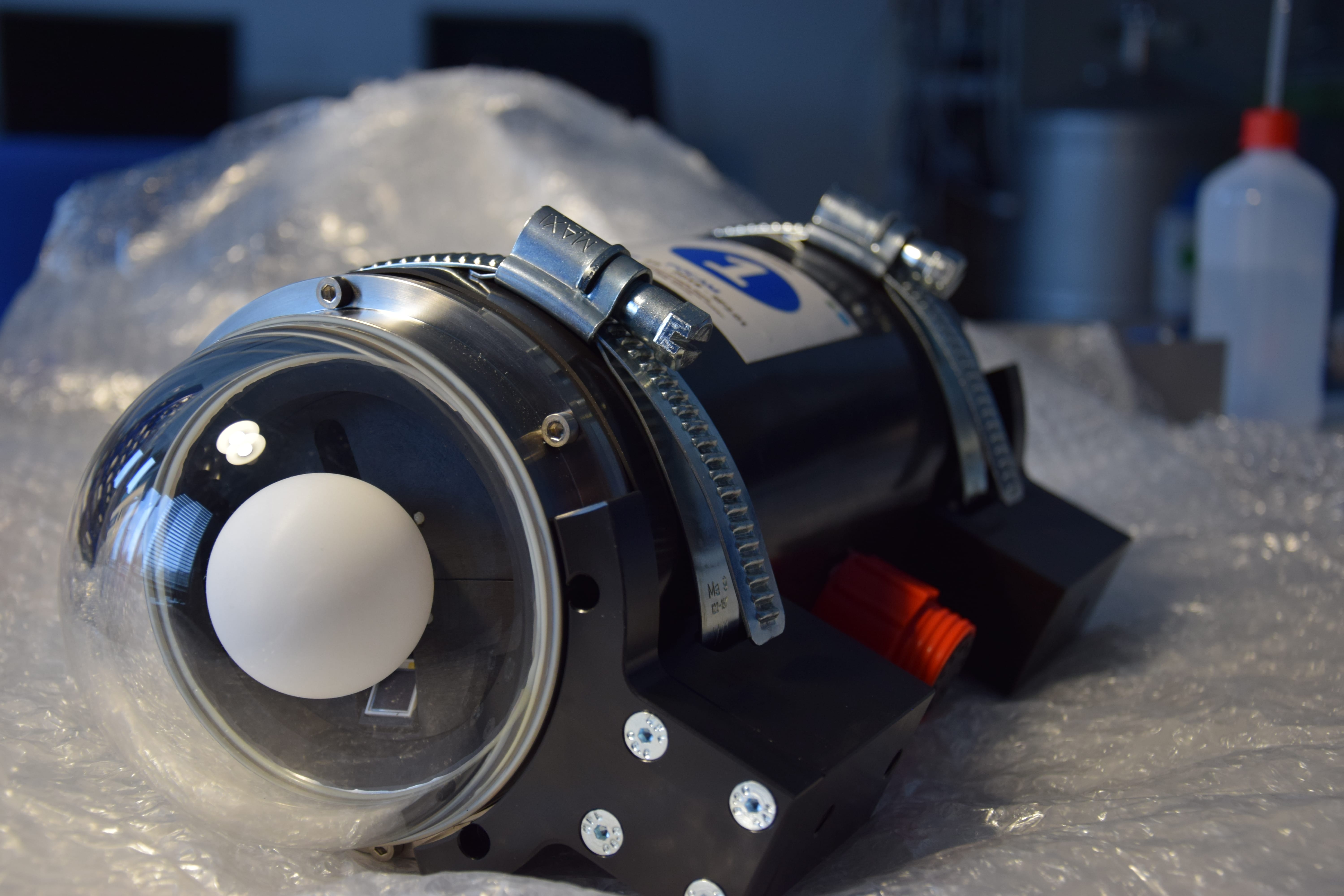}
	\caption{Finished POCAM module prior to shipment to Canada. Visible is the housing, one integrating sphere, the mounting structures and the instrument connector including its safety clamp.}
	\label{fig:pocam-finished}
\end{figure}

An end-to-end calibration procedure of the POCAM has been performed using an absolute calibrated reference photodiode. 
The reference photodiode was placed at various distances from the POCAM along its cylindrical axis facing one hemisphere. Then, the POCAM LEDs and flasher voltages were scanned and the reference calibration allowed deduction of the light output of the hemisphere together with the emission profile. This procedure was repeated for each POCAM hemisphere. An example of the calibration plot of an ADC readout versus the extrapolated number of photons is given in Fig.\,\ref{fig:pocam-calibration}. After calibration, the POCAM instruments were used for the calibration of the sDOM instruments and then made ready for shipment to Canada. The completed POCAM is shown in~Fig.\,\ref{fig:pocam-finished}.
\par
\par
\section{The STRAW Digital Optical Module (sDOM)}
\label{sec:sDOM}
The light sensor module for STRAW -- the sDOM -- is designed to register the flashes of light emitted by the POCAMs and in this way provide an estimation of the  light attenuation length at Cascadia Basin. The sDOM monitors also the background light produced by the ambient bioluminescence and radioactivity (mainly $^{40}$K). The design of the sDOM is based on the one of the POCAM, but instead of the light flasher unit, it uses two hemispherical photomultipliers.  All elements of the readout, data acquisition system and file storage are included inside the housing of the sDOM, making it an independent instrument that can communicate via Ethernet and RS485.

Also for this section, a number of plots have been adapted from \citep{Henningsen:2019jor}. More details on the sDOM characteristics, especially the integrated photosensor characterization, can be found there as well.

\subsection{Housing}
\begin{figure}[h!]
    \centering
	\includegraphics[width=0.3\textwidth]{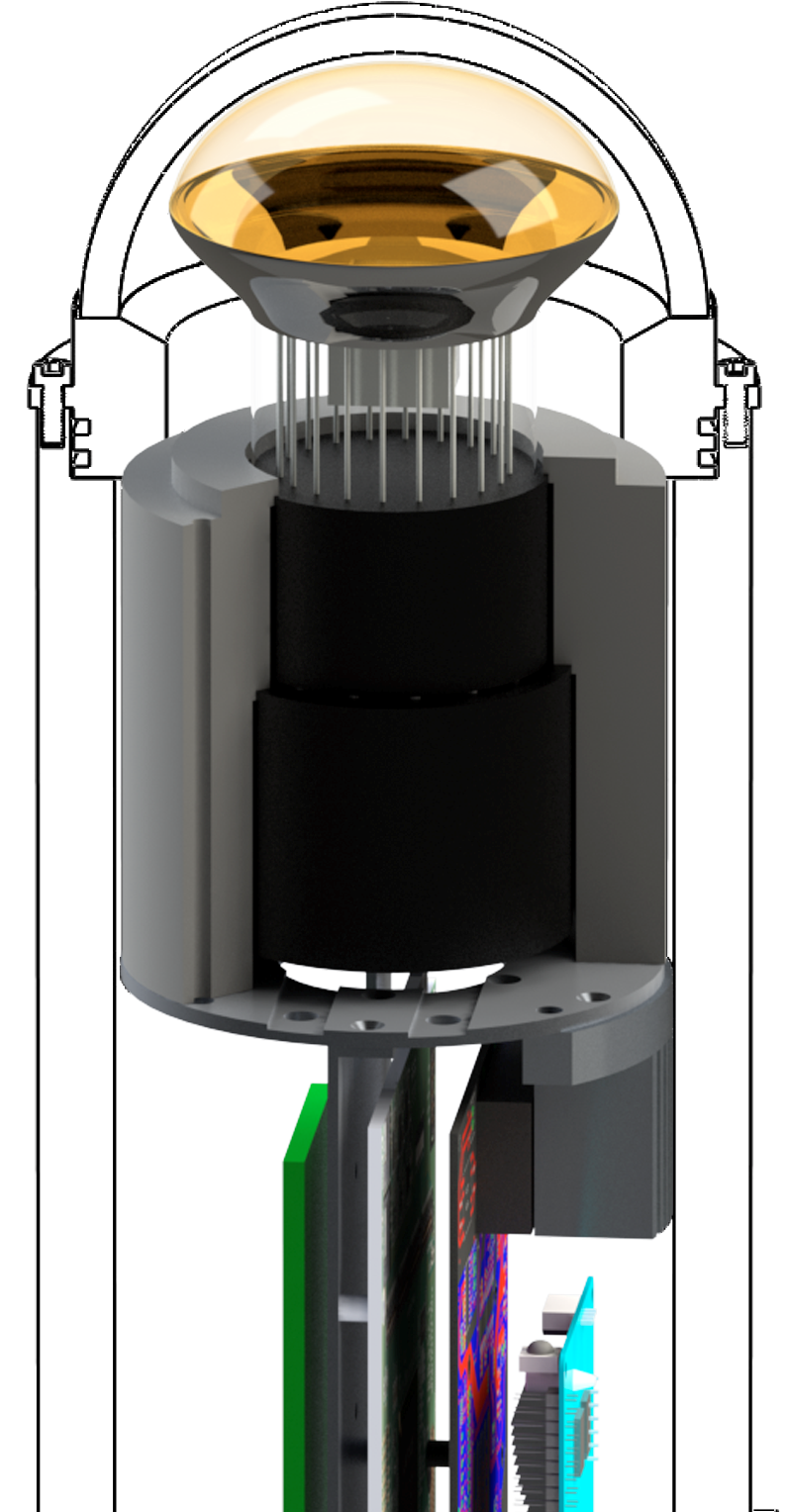}
	\caption{The interior design of the sDOM. The two PMTs are mounted in their negative forms whereas the electronic is mounted on the aluminum plate which is connected to one PMT form.}
	\label{fig:sdom-internal}
\end{figure}

%The decision for recycling the POCAM design also for the sDOM was based on the fact that it had already been successfully tested in water at great depths and also in order to reduce the number of different hardware parts. 
The housing of the sDOM uses a longer titanium cylinder compared to the POCAM one (60\,cm total length of the module) in order to allow for enough room to host the two PMTs with their HV bases, the readout electronics, and a single board DAQ computer (see~Fig.\,\ref{fig:sdom-internal}).
\par
%
%Including the two hemispheres that are identical to the POCAM ones it measures around $60\,$cm in length. 
It is designed to hold an outside pressure up to $600\,$bar, far exceeding the conditions present at Cascadia Basin.
 %Those design parameters still by far exceed those present at the Cascadia Basin.
 Additionally, the  sDOM housing was subjected to a number of environmental and stress 
 %{\color{darkgreen}(see separate section about test?) } 
 tests to ensure that the module would withstand deployment and operation at a depth of 2.6\,km b.s.l.. 
 In the sDOM housing are mounted a penetrator for power and data connections and a vacuum port for degassing. These are both identical as the one used for the POCAM.

\subsection{Photosensors}
The sDOM is equipped with two 3$''$ Hamamatsu Photonics R12199 PMTs. These are hemispherical PMTs instrumented with a Bialkali photocathode with an average 25\% peak quantum efficiency at 390 nm.  
%This spectral response provides the sDOM with the ideal optical conditions to serve efficiently all its primary tasks: monitoring of bioluminescence phenomena (whose typical emission spectrum range is 440-540\,nm), detection of the Cherenkov light emitted by the decay of the $^{40}$K present in the sea water and detection of POCAM light emission, whose LED array produces light pulses from 350\,nm to 610\,nm (see par \ref{sec:POCAM}).\\
%\begin{figure}
%\centering
%\includegraphics[width=1.1\textwidth]{img/PMT3.png}
%\caption{Schematic of PMT R12199.}
%\label{fig:3}
%\end{figure}
%
%\begin{figure}
%\centering
%\includegraphics[width=0.9\textwidth]{img/PMT1.png}
%\caption{Spectral response and gain trend of the PMT R12199.}
%\label{fig:1}
%\end{figure}
%
The HV of the PMT is provided by an active (Cockroft-Walton) base from Hamamatsu, customized for STRAW by the manufacturer (C12842-02 MOD). The sDOM is equipped with a DC/DC converter transforming the 48\,V system supply voltage  to 5\,V for the PMT. The socket uses the 5\,V power supply and an analog control input which is 1/1000 of the PMT HV.  The control voltage for the PMT (0\,V to 1.5\,V) is produced using a pulse width modulator (PWM).
%\begin{figure}
%\centering
%\includegraphics[width=0.8\textwidth]{img/PMT2.png}
%\caption{Schematic of the socket C12842-02.}
%\label{fig:2}
%\end{figure}
%
The PMTs are placed inside the two glass hemispheres.
In order to avoid reflections, refractive losses and a limited field of view, the PMTs are glued to the glass hemispheres ($n_{Glas} = 1.53$) using optical gel (Wacker \textit{SilGel 612}, $n_\mathrm{Gel} = 1.4$ at $589\,$nm) with good transmission between 350\,nm and 600\,nm. The same optical gel has been used by KM3NeT for integrating their multi-PMTs optical modules~\cite{KM3PMT}.

\subsection{Characterization of the Bare PMTs}
\begin{figure}
\centering
\includegraphics[width=1.0\textwidth]{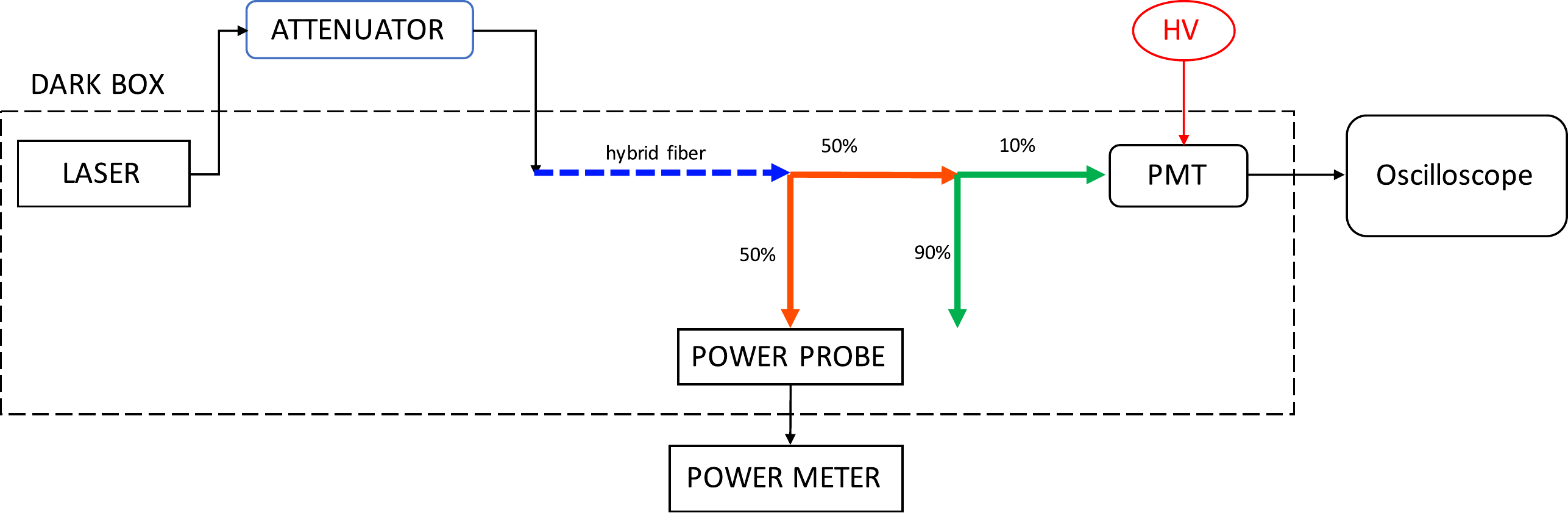}
\caption{Setup scheme for the characterization of the PMTs.}
\label{fig:13}
\end{figure}
All PMTs have been characterized and calibrated before their integration into the sDOMs. 
The following properties have been measured: the gain and saturation vs. HV, the transit time spread (TTS), and the dark rate. 
The setup used for the PMT characterization is described in Fig.\,\ref{fig:13}. It is mounted inside a sealed dark box and it is composed by: a picosecond pulsed laser (PiLas, $\lambda=(405\pm15$)\,nm, pulse width < 45\,ps, up to 100\,MHz), an optical attenuator (PiLas, attenuation range 0 to -80 dB), two different optical fiber splitters (50\% and 10-90\% emission ratio), a power meter (Newport 2936-R) with power probe (Newport 918D-UV-OD3R), and an oscilloscope (Teledyne LeCroy HDO6054, 500\,MHz, 2.5\,GS/s). 

\begin{figure}
\centering
\includegraphics[width=\textwidth]{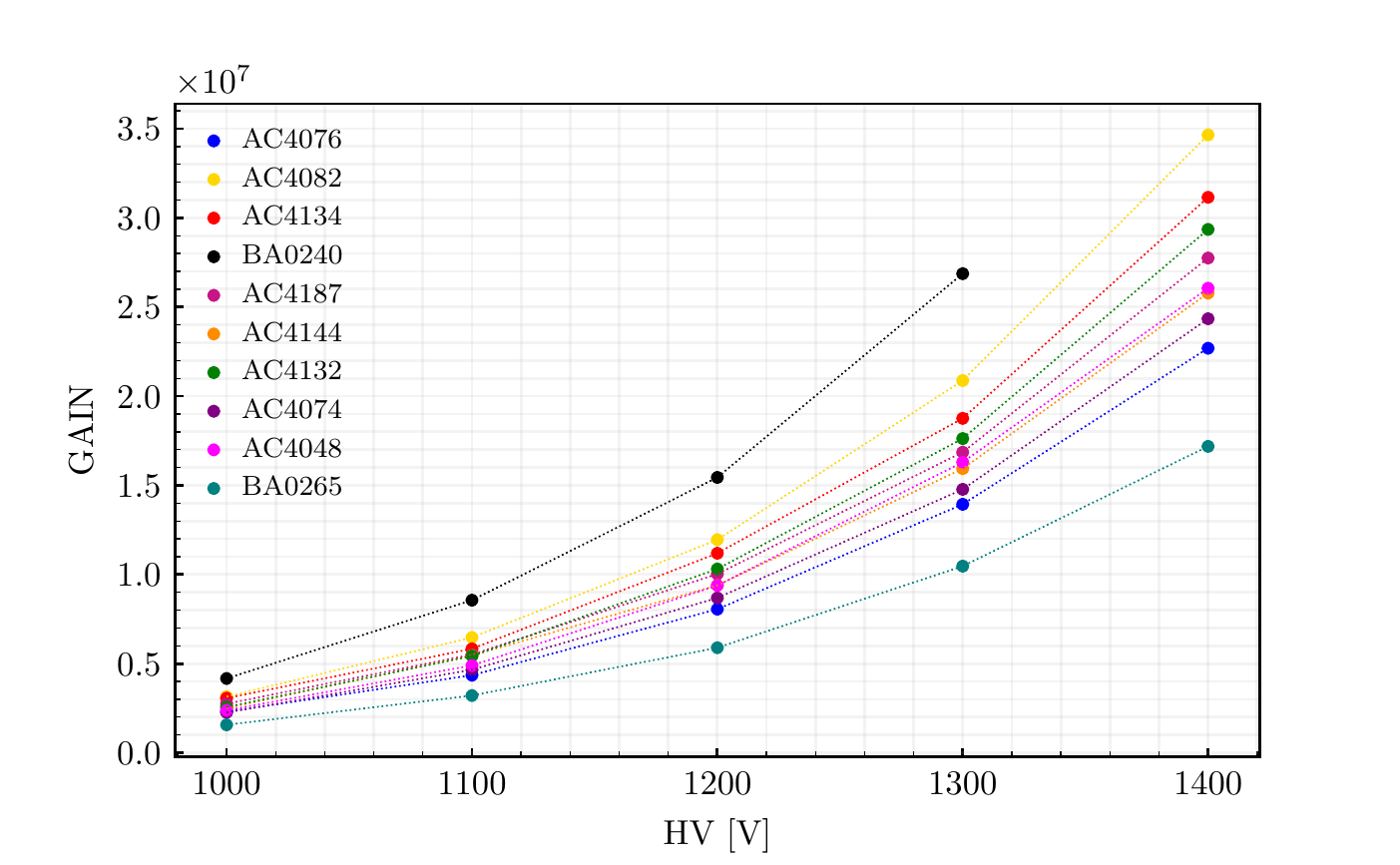}
\caption{Gain vs. supply voltage (set with the socket) for every PMT of STRAW. \rev{For the BA0240 the gain has been measured only up to 1300\,V as it already reached a value $>2.5 \cdot 10^7$ fo that voltage}.}
\label{fig:14}
\end{figure}

In order to measure the gain of the PMT, the single photoelectron charge has been determined at different HV values (see Fig.\,\ref{fig:14}). 
%this has been done after scaling down to the single photon level the laser light emission.  
Simultaneously, for each HV value, the amplitudes of the single photoelectron signals have been measured with the aim of identifying the trigger thresholds for next steps.
Each PMT has been tested with its own socket: the high voltage power supply of the PMT is enabled after setting the control voltage of the socket. \\
\rev{Also long-term monitoring of the dark rate trend was} performed revealing a day-night modulation induced by a slight temperature change in the laboratory. %(Fig.\,\ref{fig:6}).%
This modulation is no matter of concern since in its final working conditions after the deployment at 2600 m depth in the Pacific Ocean, STRAW will operate at a (stable) temperature of about 2$^{\circ}$C.

%With the aim of discriminating the background produced by bioluminescence and $^{40}$K decay, the dark noise of every PMT has been measured. A long-term monitoring of the dark rate trend revealed the day-night modulation reported in fig \ref{fig:6}.
%It is worth remarking that the measured values of the dark noise (taken at room temperature) can be considered as an upper limit. Indeed, in its final working conditions after the deployment at 2600 m depth in the Pacific Ocean, STRAW will operate at a (stable) temperature of few degrees.
%Anyway, even the fluctuation of dark noise produced by a variation of the ambient temperature of few degrees can be considered negligible compared to the rate measured after coupling the PMT with the glass hemisphere, fig \ref{fig:7}. In this case, the large increase of dark noise is caused by the radioactivity of the glass envelope [ref XXX].
%

%\begin{figure}
%\centering
%\includegraphics[width=\textwidth]{img/sdom/darkrate_temp.pdf}
%\caption{PMT dark rate trend compared to the temperature trend.}
%\label{fig:6}
%\end{figure}
%

%\begin{figure}
%\centering
%\includegraphics[width=0.8\textwidth]{img/sdom/DArk_rate_PMT_gel.pdf}
%\caption{Dark rate measured after the integration of the PMT in the %SDOM, with the gel and the glass hemisphere.}
%\label{fig:7}
%\end{figure}

\begin{figure}
\centering
\includegraphics[width=0.9\textwidth]{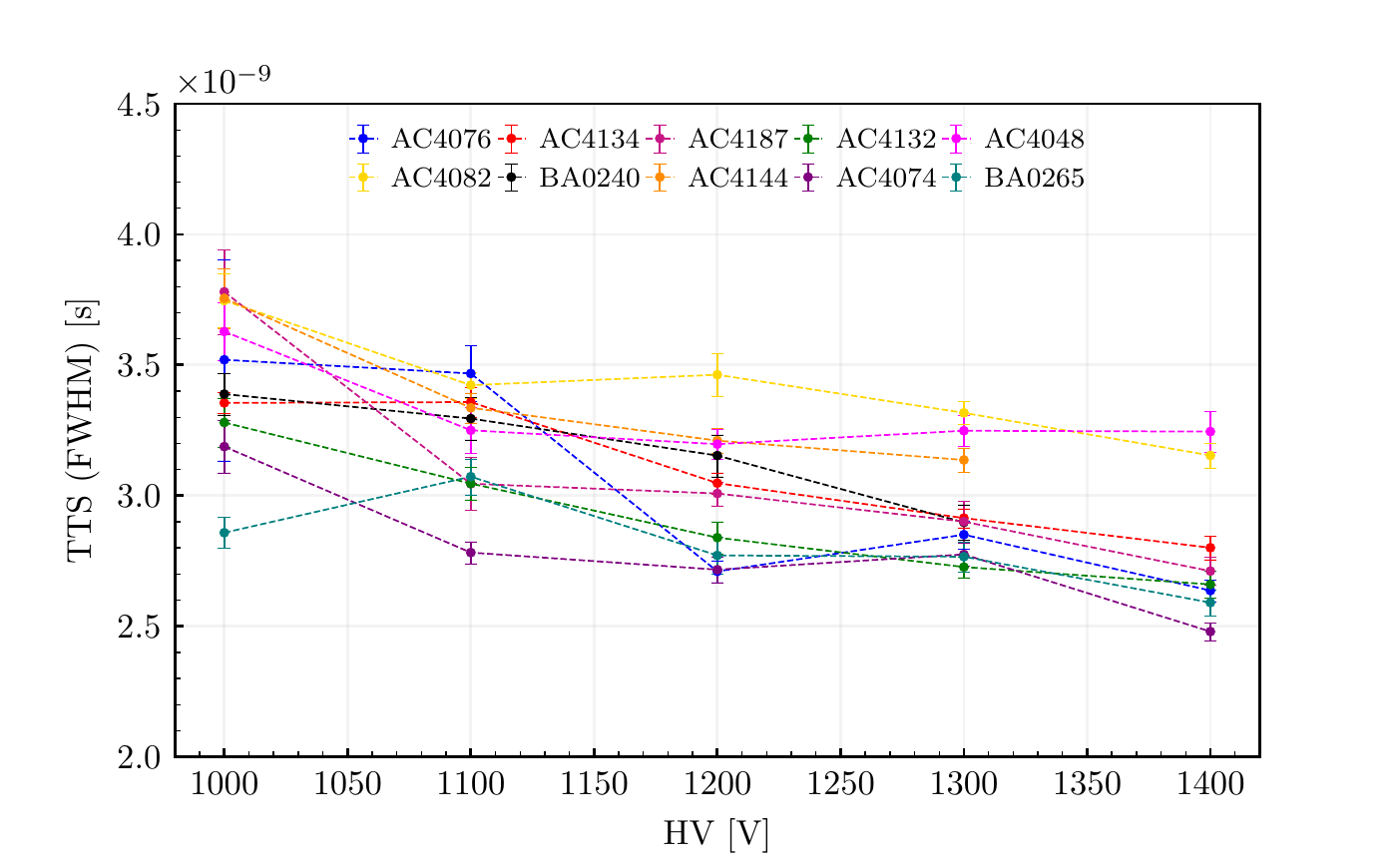}
\caption{Transit time spread trend at different HV per every PMT of STRAW: the value decreases with the increasing power supply.}
\label{fig:8}
\end{figure}

\begin{figure}[h!]
\centering
\includegraphics[width=\textwidth]{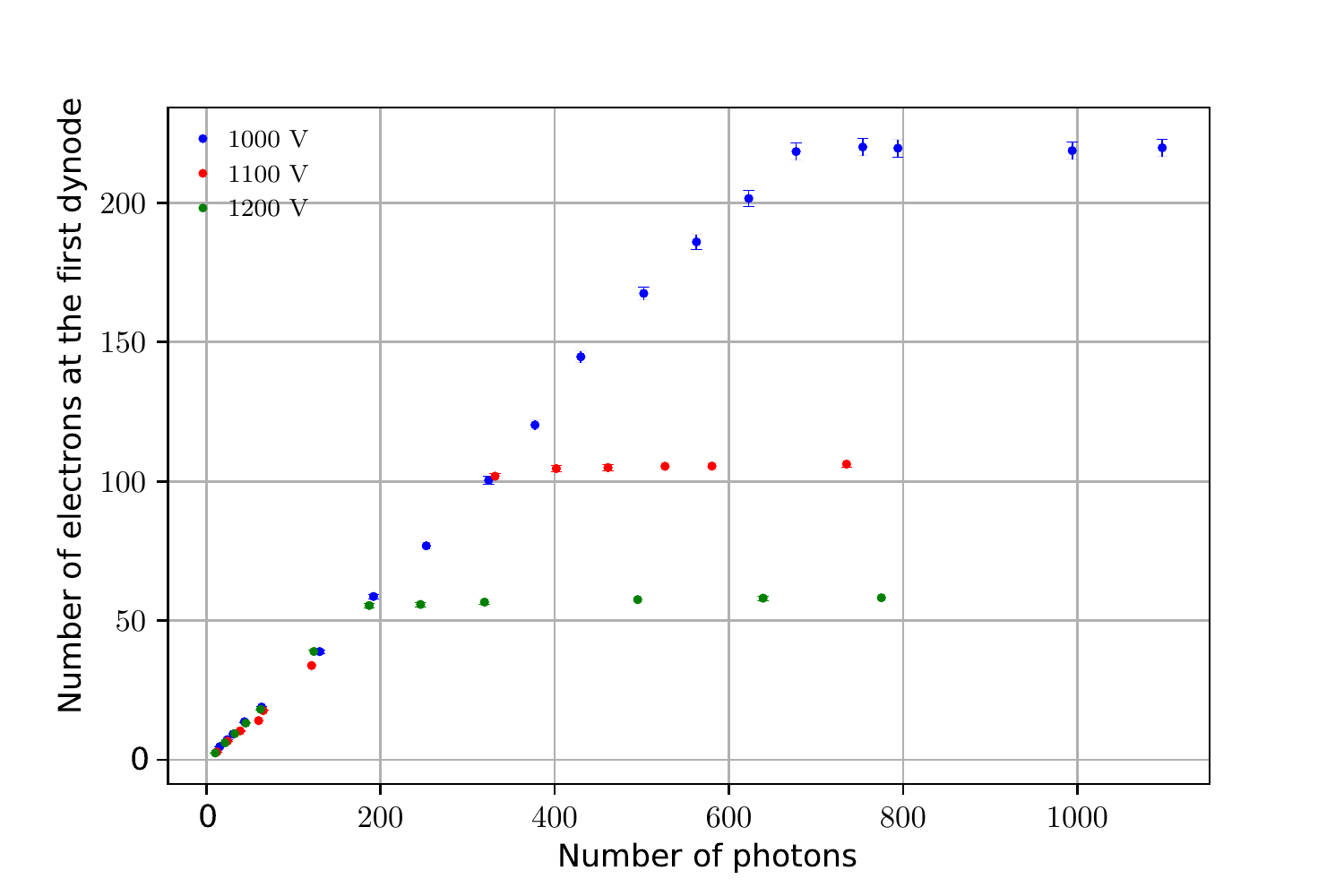}
\caption{Saturation \rev{at different values of HV.} \rev{Anode charge divided by the gain in units of elementary charge} ($e^-$ collected at the first dynode) vs. the number of photons \rev{impinging on} the PMT. The slope of the straight line in the linearity region is the quantum efficiency of the photocathode. The error bars \rev{take into account statistical errors only (systematic errors might exceed those but are not considered)}.}
\label{fig:9}
\end{figure}

The Transit Time Spread (TTS) is the intrinsic fluctuation of the transit time of a photoelectron within the PMT.  We have measured it with the setup described above  (example in Fig.\,\ref{fig:8}). We found that our measurements are in agreement with the TTS measurements done by the KM3NeT collaboration on the same type of PMTs \cite{KM3PMT}.
%When a photocathode is fully illuminated with single photons, the transit time of each photoelectron pulse has a fluctuation \cite{HamamatsuHandbook}. Estimating this fluctuation, called Transit Time Spread (TTS), that has been measured at different high voltage values, is fundamental because it impacts on the determination of light pulses arrival time. 
%In order to illuminate completely the photocathode with single photons, the distance to set between the PMT and the light output of the optical fiber, has been calculated considering the numerical aperture of the fiber and the size of the photocathode diameter.
%Once in the single photon condition, triggering on the laser output, has been measured the time interval between the laser output (START) and the peak of amplitude of the PMT signal (STOP) \cite{Vivolo:2014jma}: this time is the sum of the Electron Transit Time (ETT) and the time that light needs to propagate through the fibers and splitters chain. Since this propagation time is fixed, it has not influenced the TTS measurement: time values have been acquired by the oscilloscope at different power supply for the PMT and the histograms of the Gaussian distributions has been analyzed.\\
%The TTS has been calculated as the FWHM of these distributions (that corresponds to 2,35$\sigma$) and the results (example in Fig. \ref{fig:8}) are fully in agreement with the TTS measurements done by KM3Net Collaboration on the same PMT type \cite{KM3PMT}.\\
%
The study of the linearity and saturation of the PMT was also carried out and fine-tuned in order to estimate the amount of POCAM light needed to reach a certain sDOM over the two strings but without saturating them.
%shows a good linearity in output current over a large range of incident light level but if the incident light amount is too large, the output signal begins to deviate from the ideal linearity. 
%In order to define this range in terms of detected photons, and consequently predict the amount of POCAM light needed to reach a certain sDOM over the two strings, the saturation behaviour of the PMT has been investigated: the PMT has been illuminated with different amount of light, monitored constantly with the power meter, and powered with different HV values. 
In Fig.\,\ref{fig:9} we show the trend for the saturation measurement of one of the PMT mounted in an sDOM.
%: according to predictions, since the GAIN increase with the increasing power supply, at the same amount of light (i.e. same amount of photoelectrons produced by photocathode), the dynodes chain produces a greater number of charges on the anode, that start to deviates from its linear behaviour.
The number of photons has been calculated taking into account the readings of the power meter values, the frequency and the wavelength of laser light emission, while the number of electrons at the first dynode $N_{e^-}$ is calculated according to:
\begin{equation}
	N_{e^-} = \frac{Q_{TOT}}{G \cdot q_e}
	\label{saturation}
\end{equation}
where $G$ is the Gain previously measured, $Q_{TOT}$ is the total charge on the anode measured with the oscilloscope, and $q_e$ is the electron charge. 
\rev{In the linearity region, the slope, which is the ratio between the number of photoelectrons that reach the first dynode and the number of photons that hit the photocathode, is a product of the quantum efficiency and the collection efficiency of the PMT. From Fig.\,\ref{fig:9} it can be derived that the three slopes are in agreement and point to a value for the quantum efficiency (QE) of the PMT of around 29\%.} 

\subsection{Characterization of the Integrated PMTs}
In addition to the characterization of the bare PMT, possible effects on the final performances of the sDOM introduced by the housing and the optical gel were also investigated. In particular, we performed a  temperature-dependent characterization of the dark rate of the sDOM.  
%This should allow later to quantify effects induced by ambient radioactivity and biolumenescence. 

The sDOM flanges containing the PMTs coupled through the optical gel to the encapsulating glass, were placed in a temperature controlled environment, in which the temperature was modified in steps between 0$^\circ$C and 25$^\circ$C.  The PMT output was then monitored using a $350\,$MHz-bandwidth digital oscilloscope (Picoscope 6403D) and a sampling interval of $1.6\,$ns. A pulse-detection algorithm was then used to measure the distance in time between subsequent noise pulses ($\Delta t$) and in this way monitor the dark spectra of all sDOM PMTs.

\begin{figure}[h!]
\centering
\includegraphics[width=\textwidth]{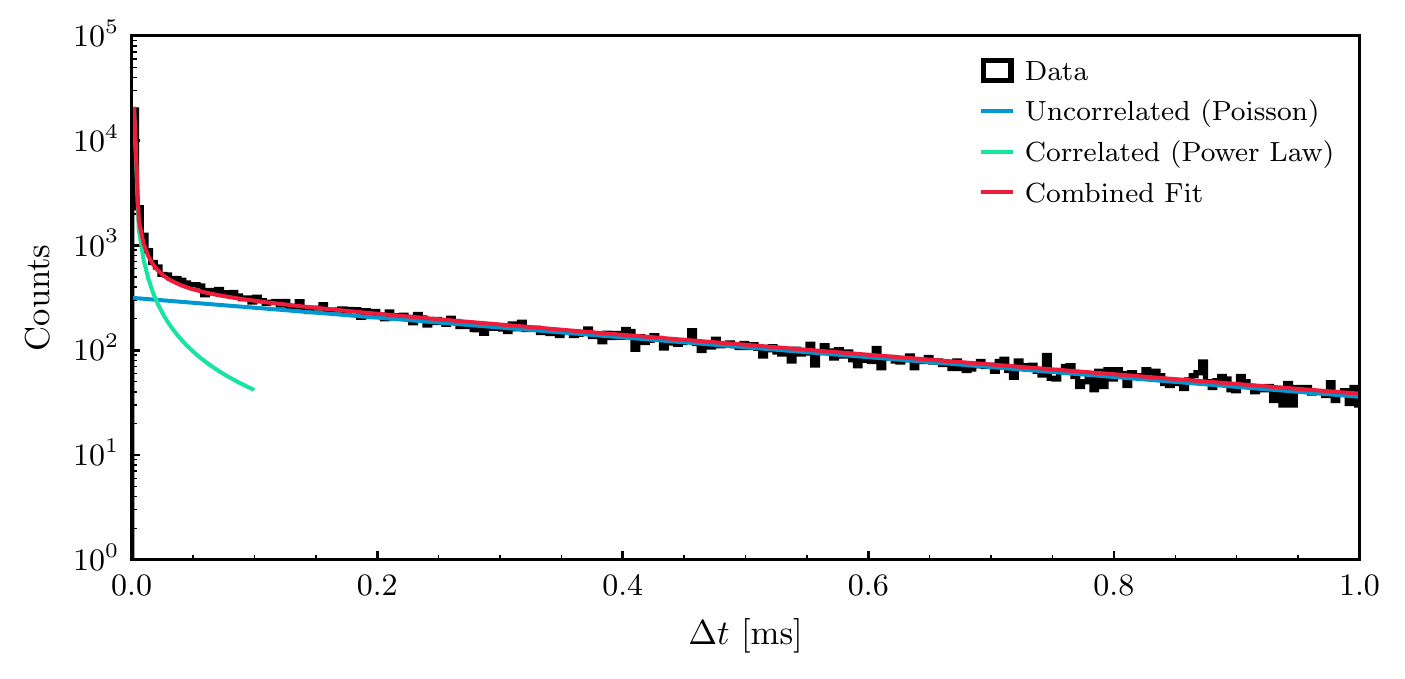}
\caption{Exemplary $\Delta t$ \rev{distribution} for pulses recorded for one of the sDOM PMTs. Shown are the noise components from uncorrelated sources, i.e~thermal and radioactive emission as well as correlated noise coming from various sources. \rev{Fits are also shown}.}
\label{fig:pico-dt}
\end{figure}

\begin{figure}[h!]
\centering
\includegraphics[width=\textwidth]{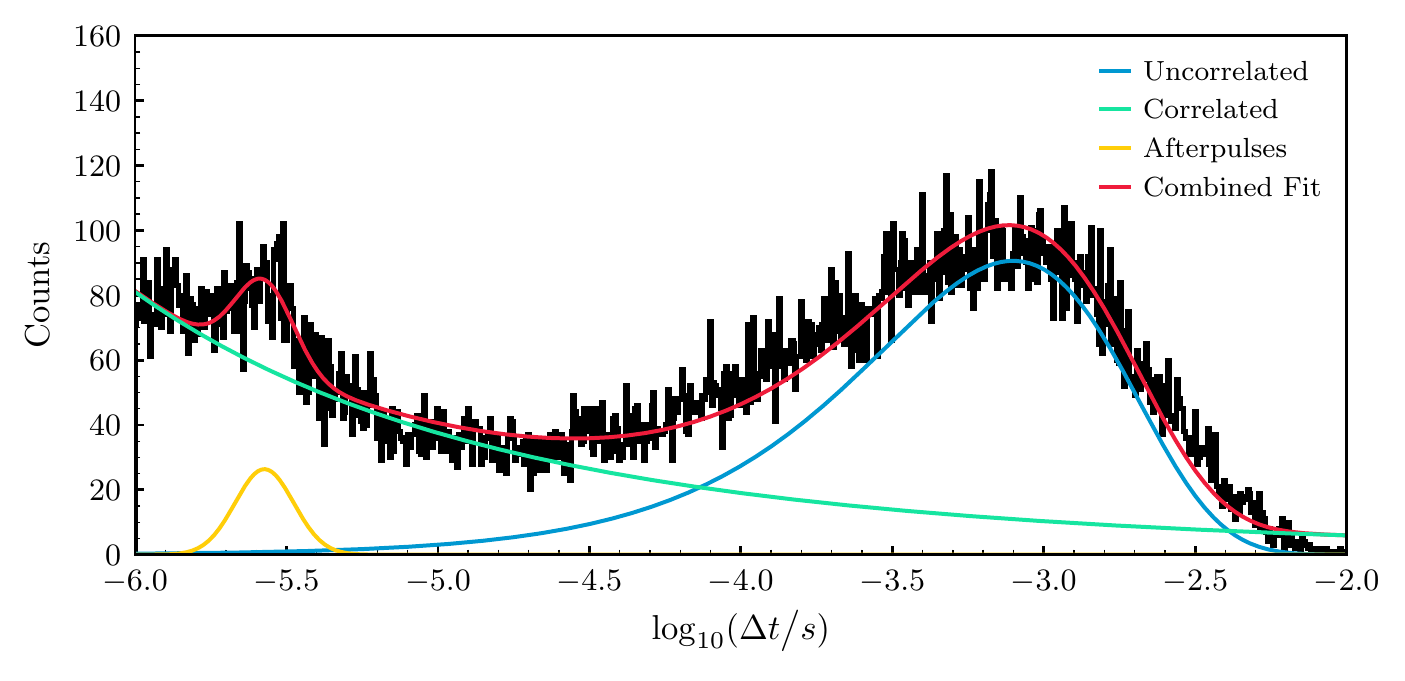}
\caption{Exemplary $\Delta t$ \rev{distribution} given Fig.\,\ref{fig:pico-dt}. Plotting the data like this more clearly shows different noise components present in the PMT.}
\label{fig:pico-log10dt}
\end{figure}

\begin{figure}[h!]
\centering
\includegraphics[width=\textwidth]{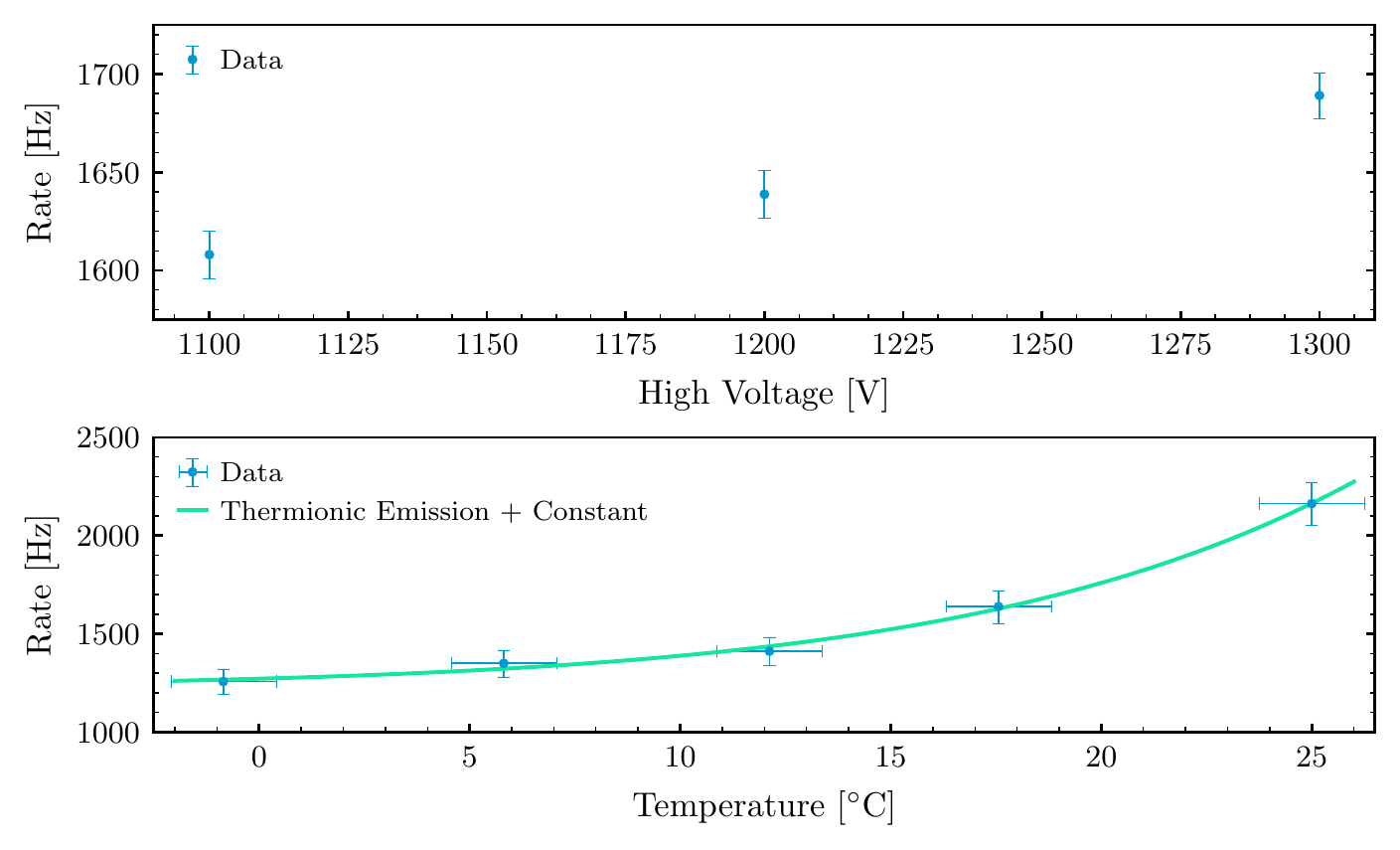}
\caption{\rev{Dark-rate of an exemplary sDOM PMT as function of HV (upper plot) and temperature (lower plot). The solid line in the lower plot represents a fit to the data consisting of a simple model for the thermionic emission ($\propto T^2 \exp(-W / kT)$, Richardson Law) and a constant offset, which most likely is caused by ambient radioactivity.}
%Using the known thermal emission of PMTs and the measurement of raw data at various temperatures, a baseline calibration was performed. The top plot shows the uncorrelated noise rate versus voltage, the lower plot a fit of the latter including a constant offset induced by ambient radioactivity.
}
\label{fig:pico-temp-volt}
\end{figure}

From the spectra in  Fig.\,\ref{fig:pico-dt} and Fig.\,\ref{fig:pico-log10dt} three components are clearly visible: the uncorrelated noise, the correlated one, and the afterpulses. The uncorrelated noise follows the expected Poissonian distribution and originates from thermionic emission of the PMT and the ambient radioactivity. This manifests itself as an ordinary exponential and offers a measure of the mean dark-rate of an sDOM hemisphere. The correlated noise, on the other hand, deviates from the Poissonian noise at low-$\Delta t$. It is much less well understood although it is most commonly attributed to bursts of photons due to scintillation and Cherenkov radiation in the encapsulating glass and the optical gel~\cite{unland:thesis:2017, stanisha:thesis:2017}.
However, using the Richardson Law~\cite{richardson} for thermionic emission one is able to characterize the uncorrelated noise, i.e. the mean darkrate, as a function of temperature and voltage. The latter is shown in Fig.\,\ref{fig:pico-temp-volt} and has been carried out for each sDOM hemisphere. \rev{At room temperature and with HVs between 1.2\,kV and 1.3\,kV dark-rates of the order of 2\,kHz were measured for the glued-in PMT. We did not characterize the after-pulsing of the PMTs in detail as we do not expect this to impact our measurement.}

\subsection{Readout Electronics}
\begin{figure}[h!]
\centering
\includegraphics[width=\textwidth]{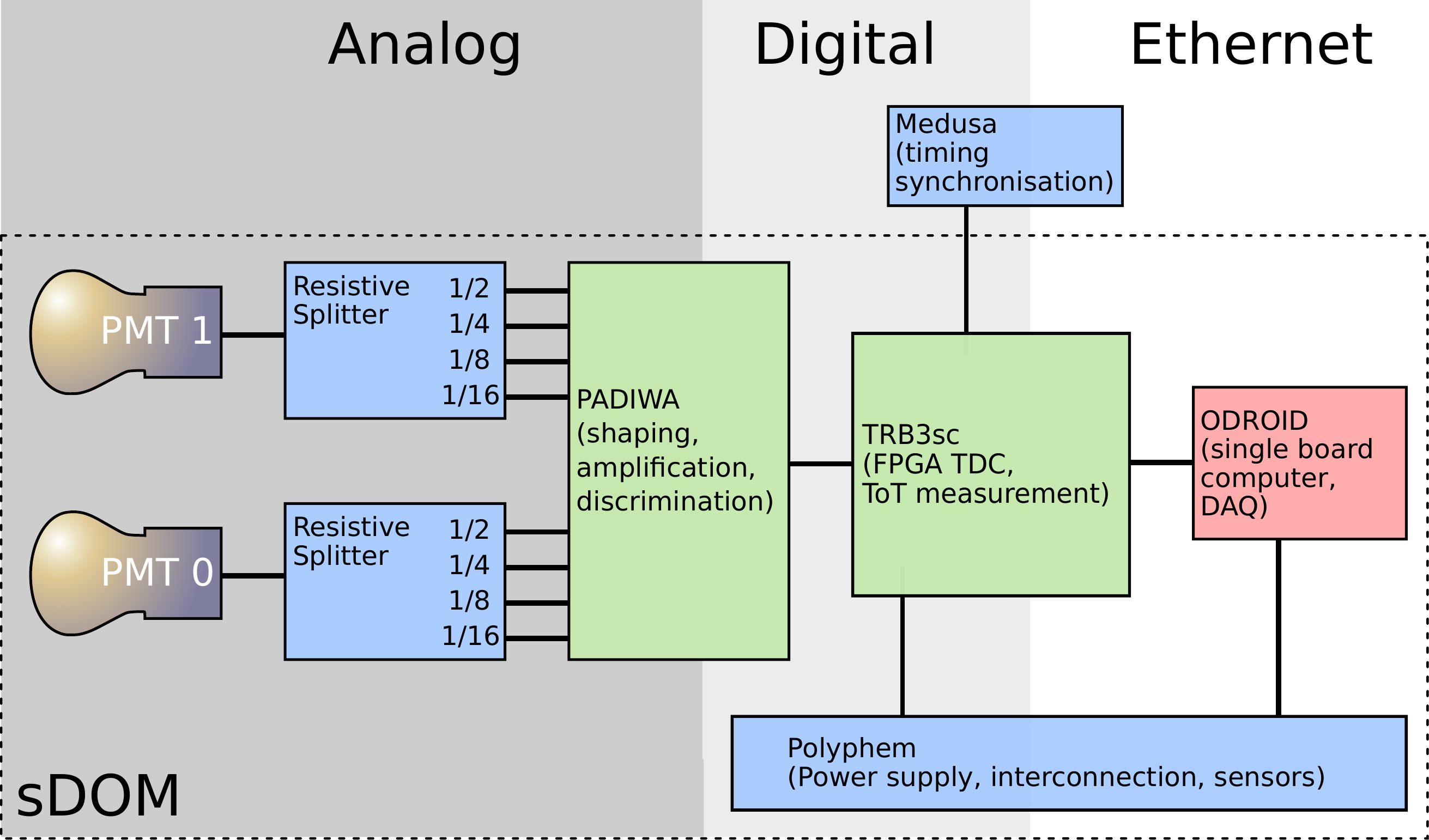}
\caption{\rev{Schematic drawing of the sDOM electronics. Custom designed and produced electronic parts are shown in blue, TRB hardware by GSI is shown in green, and the Odroid C2, which could be bought from stock is shown in red.}}
\label{fig:sdom_scheme}
\end{figure}
The sDOM electronics was designed as a four-layer stack of printed circuit boards. This allowed integrating existing boards into the special geometry of sDOM housing, keeping development efforts at a minimum, while achieving a mechanically reliable and robust setup. \rev{A schematic overview of the single elements can be found in Fig.\,\ref{fig:sdom_scheme}.}
%{\color{red} Two pictures of setup, showing the mechanical setup; side view and top view}

The main control of sDOM is taken over by an Odroid C2 computer \cite{odroidc2} running on a standard Armbian OS \cite{armbian} with few minor patches to allow access on the serial console port. The internal GbE ethernet port is used to connect to the TDC board, and a standard USB ethernet adapter handles the connection to the surface (for more details, refer to \cite{Henningsen:2019jor}).
A modified U-boot inside the Odroid allows console access via slow RS232 connection even in case of a broken eMMC storage device. For that case, a spare micro SD card can be used to reinstall the Armbian OS.

An auxiliary board (aka Polyphem) acts as a connector board between the Odroid C2 and the FPGA board. This auxiliary board handles all connections to the surface by implementing galvanic insulation for all signals, as well as converting the incoming 48\,VDC power to the internal 5\,VDC supply voltage. Power to different subsystems can be switched here, allowing the system to start up automatically with minimum power requirements. Two additional DS18B20 temperature sensors are mounted on the PMT bases. Besides, two PWM DACs allow setting the PMT HVs by simple Linux commands.

Besides, several sensors are implemented: a DS1822 temperature sensor provides the DC/DC stage temperature (and providing unique 48\,bit ID for all sDOMs). An MS8607 sensor provides temperature, relative humidity and pressure inside the sDOM housing for long term monitoring. For orientation sensing an MV6470 sensor provides acceleration and magnetic field data.
All sensors can be read out directly from Linux.

Measurement of PMT signals is handled by a standard \textit{Trigger Read-out Board} (TRB3sc), in connection with a Padiwa frontend card, both developed by GSI in Darmstadt. \rev{The TRB3sc is used for recording time over threshold (ToT), including precise time stamps, of the PMT pulses at four different voltage levels, which allows for a rough charge reconstruction}. These boards use the TRBnet protocol \cite{trbnet}, have been successfully used in several experiments and provide a complete package of readout and analysis via the DABC \cite{dabc} package. The TRB3sc operates in stand-alone mode, implementing the whole DAQ system (CTS and TDC) in one FPGA under control of standard GbE UDP. The DAQ can be controlled and read out via Ethernet, using \rev{an SSH connection}; for simplicity during the test phase, a VNC server allows full control of the whole sDOM infrastructure.

\begin{figure}[h!]
\centering
\includegraphics[width=\textwidth]{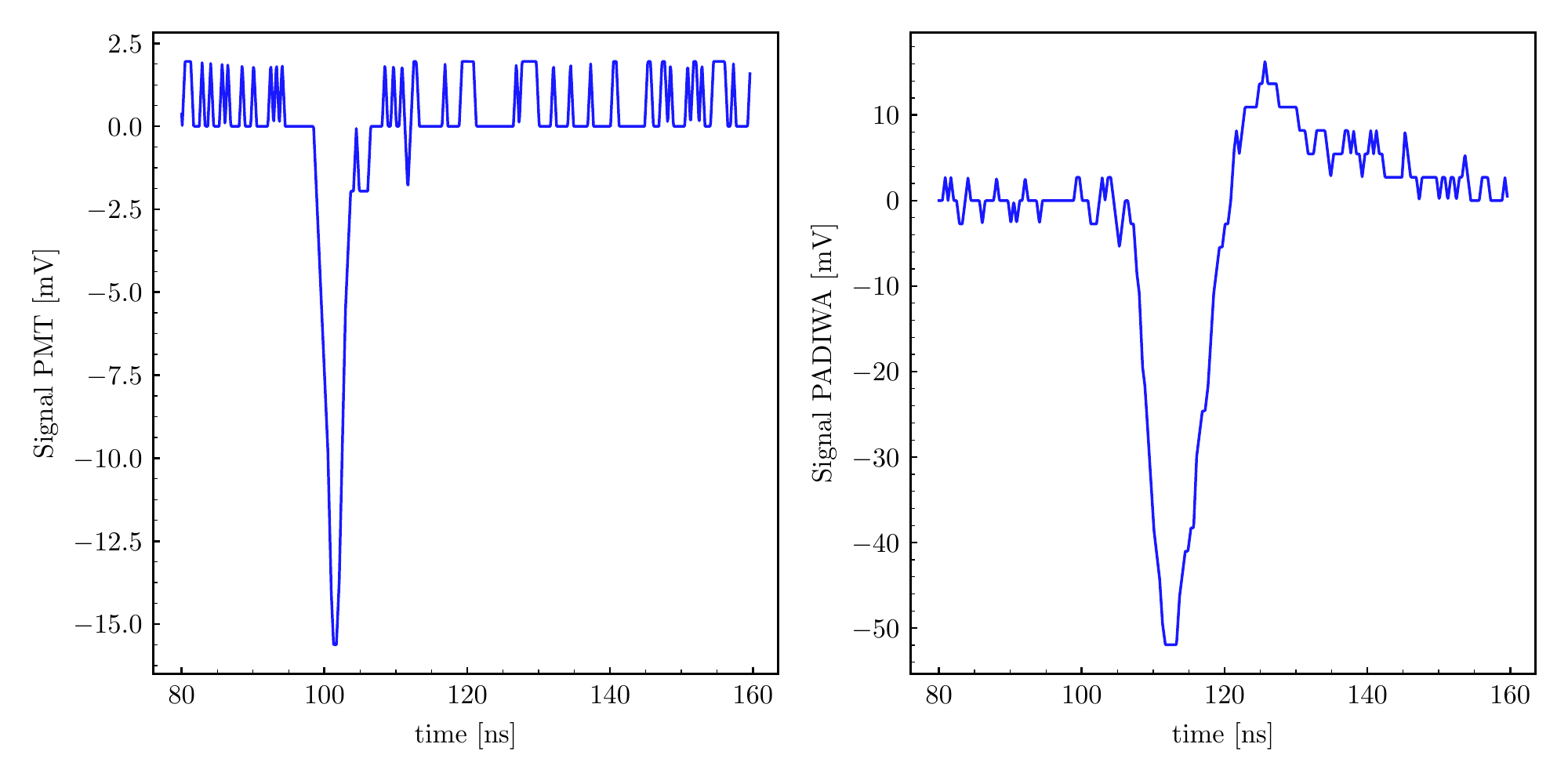}
\caption{A typical single photo-electron pulse signal at the output of the PMT and after the Padiwa amplifier before the discriminator. \rev{The data were recorded using a USB digital oscilloscope (PicoScope 6403D).}}
\label{fig:SPE_pulse}
\end{figure}

Each PMT is connected by a four channel resistive attenuator board to the Padiwa. This allows measuring timing of PMT signals at different levels. The pulse shape of PMT signals can therefore be easily reconstructed and should provide better pulse charge information than a simple one channel ToT measurement. Fig.\,\ref{fig:SPE_pulse} shows an exemplary PMT pulse before the resistive attenuator and after amplification and shaping by the Padiwa, \rev{as it enters the discriminator}. The data have been recorded simultaneously using a USB digital oscilloscope (PicoScope 6403D).

To make use of the high precision TDC measurement on the 10\,ps scale\,\cite{Neiser:2013yma}, a central timing distribution had to be implemented in STRAW. This system is implemented inside the mini junction boxes and described later (see Sec.\,\ref{sec:anchor}).
Due to copper cabling limitations, the \rev{standard} TRBnet functionality for timing distribution could not be used.
%In an upcoming STRAW-B setup this limitation will not be present anymore.
%}

\subsection{Internal Mounting \& Heat Dissipation}
The design of the sDOM housing is based on the precursor design of the  POCAM one. As such, only the length of the cylinder was adapted to fit the necessary electronics and various other components necessary for full functionality. The PMTs are encapsulated in polyoxymethylene (POM) holding structures whereas two \rev{leaf springs} ensure leeway against thermal and mechanical stress. The internal structures are fixed to the housing using four threaded bars. A central aluminum plate \rev{acts} as mounting point for the electrical backbone (see~Fig.\,\ref{fig:sdom-internal}).

For the heat dissipation, two \rev{primary} heat sources had to be considered: The FPGA which is placed on the TRB3sc and a voltage converter on the power supply board. Since there is little air flow in the module, the cooling must be established by a connection to the titanium housing. 
\rev{Both components are coupled by thermally conductive pads to the central aluminum plate, which in turn is thermally coupled to the titanium housing.}
%\rev{By means of an electrically isolating thermal pad the FPGA is coupled to the central aluminum plate, which in turn is thermally coupled to the housing}.
%As for the voltage converter heat sink, an aluminum L-profile applies pressure to a heat conductive pad on the voltage converter, thus coupling it via the aluminum plate indirectly to the outside of the instrument.
%While in-air operation was possible, temperatures were reaching critical levels after about an hour of operation. 
Tests in $17\,^\circ$C water show stable temperatures \rev{of the PMTs of less than $2^\circ$C above the temperature of the ambient water} over various hours of operation confirming a sufficient heat dissipation of the sDOM. 
%For more details we refer to the master thesis in ~\cite{Henningsen:2019jor}.

\section{The Strings}
\label{sec:string}
As introduced in Sec.\,\ref{sec:intro}, STRAW is composed by two strings (mooring lines) named yellow and blue for practical reasons. The yellow string is equipped with two POCAMs and two sDOMs whereas the blue string has one POCAM and three sDOMs. In order to cover \rev{different baselines} the modules are mounted at different heights of 30, 50, 70 and 110\,m above the seafloor. 

A two-line holding structure \rev{on which the instruments are mounted ensures that the strings do not twist by more than what still allows for measuring with an unobstructed light path between the instruments. Spacers put between the steel cables every 5\,m ensure a restoring torque proportional to the line twist, which is caused by the pull of the buoyancy. Assuming a water current of less than 20\,cm/s, we estimated a total line twist of less than 30 degrees.}
The cables are aligned next to each other, and they are fixed within the module mountings installed every five meters. Stronger top and bottom spacers serve as connection to the anchor and the buoy at a height of 146.5\,m. A swivel between the top spacer and the buoy is added as an additional degree of freedom. In this way, the buoy can freely rotate and does not affect the string itself.

To preserve the mechanical parts from the saltwater, all steel parts are hot-dip galvanized and additionally coated with a black polyurethane based color for sub-sea applications (SikaCor EG-5). The black coverage also reduces the reflections on the metal surface. 

\section{Seafloor Infrastructure and Anchor Design}
\label{sec:anchor}
The NEPTUNE observatory of Ocean Networks Canada consists of an 840\,km fiber optical cable loop with several nodes in the northeast Pacific Ocean. From the node in Cascadia Basin, \rev{an oil filled sub-sea cable (manufactured by Teledyne ODI)} connects to a junction box, which acts as a distribution point for various experiments. Each of the two STRAW mooring lines is equipped with an ONC-provided mini junction box at the bottom.
The mini junction box provides an intermediary between the main junction box and the modules. 
The connections from the STRAW instruments to the mini junction box were made on the ship, the connection from the mini junction box to the junction box was made under water by the ROPOS remotely operated underwater vehicle (ROV).

Each mini junction box contains (besides the ONC infrastructure for power and network) a special synchronization board (named Medusa). A microcontroller handles Ethernet accesses (either web-based, or UDP), allowing either script-based or web-controlled setup of the synchronization signals. An FPGA provides sync signals, based on a high precision TXCO oscillator, and distributes the pulses over an LVPECL clock distribution chip. \rev{The individual cable delays have been characterized in the lab with a precision of less than 10\,ns. A more precise evaluation is still ongoing. Delays inside the sDOM PMTs and electronics are assumed to be similar enough and are therefore neglected. Additional timing uncertainties are added by uncertainties in the geometry (mainly the tilt of the strings).}

Special precautions in cabling allow to operate the strings from either mini junction box, or even in standalone mode in case of failures. To compensate for potential cabling issues, all signals leaving the mini junction box to the sDOMs (i.e., GND and differential sync) can be disconnected by relays. Together with galvanic input insulation on the sDOM side, broken cables can be completely disabled.

Using the length of the sync signals, the \textit{TRB3sc} internal FPGA counters can be reset synchronously. By using the measured cable delays, TDC hits can be analyzed with a common time in both strings.

%The anchor is composed by a dead weight of two train wheels (800\,kg) coupled to a rotating structure. This structure allows the {\it in-situ} adjustment of the relative angular alignment of the strings in order to have a free line of sight between the modules mounted on the two mooring lines of STRAW. During the deployment the anchor is fixed with a solid pin to avoid the structure from rotating in an uncontrolled way. With the mooring lines on the seafloor, the ROV removes the pin and disengage the rotating plate from the train wheels. The angular adjustment is done by the  ROV turning the strings by pulling on a tow with monkey fists on the end. Once in position the structure is again fixed with the solid pin.
%Several sacrificial anodes are additionally attached on the anchor to  protect the critical steel parts from damage due to electro-chemical reactions.

\begin{figure}[h!]
    \centering
	\includegraphics[width=0.6\textwidth]{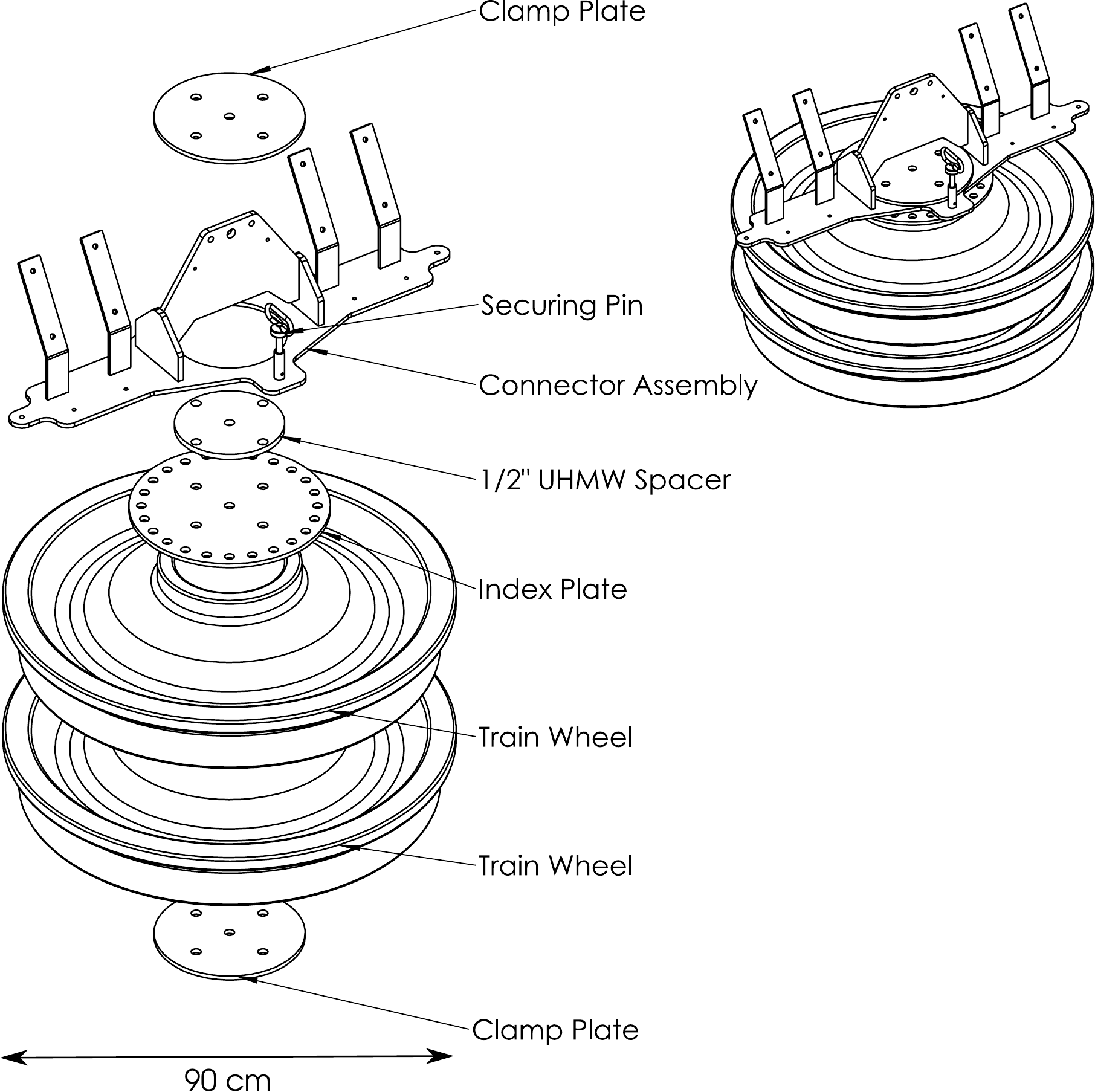}
	\caption{Anchor design by Dirk Brussow of ONC. The rotating structure is held together by threaded bars connecting the clamp plates. The two steel lines of a string are connected to the connector assembly. The connector assembly part also provides two mounting points for two ODI connector plates.}
	\label{fig:anchor}
\end{figure}

The anchor in Fig.\,\ref{fig:anchor} is \rev{composed two train wheels that act as dead weight (625\,kg weight in water), coupled to a structure which allows rotating the strings after deployment and hosts connector assembly for sliding in the ODI wet-mate connectors.} 
%This way a free line of sight between the modules mounted on the two mooring lines of STRAW could be ensured. 
\rev{This way both strings can be aligned to ensure a free line of sight between all modules.} During \rev{the deployment,} this structure on the anchor is fixed with a securing pin to avoid the structure from rotating in an uncontrolled way. With the mooring lines on the seafloor, the ROV removes the pin and disengages a rotating plate from the train wheels.
\rev{Now the ROV adjusts the alignment of the string by pulling on a tow with a monkey fist on the end, which is attached to the rotating plate.}
%The angular adjustment is done by the ROV turning the strings by pulling on a tow with a monkey fist on the end. 
Once in position, the structure is again fixed with the securing pin.
\rev{The anchor and steel lines are protected from corrosion by sacrificial anodes attached to the anchor. Besides, the rotating structure is painted with a zinc-based paint.}
%The rotating structure on the anchor is protected from corrosion by sacrificial anodes and a zinc-based paint.

%
\par
%

%{\color{green} Therefore it is possible to rotate the holding structure for the strings on the dead weight of two train wheels (800kg) and fix it in 15 degree intervals. During the deployment the anchor is fixed with a solid pin to save the structure from rotating. Once the strings are on the seafloor the ROV operated from the vessel is able to pull the pin out and disengage the rotating plate from the train wheels. After this the ROV is capable of turning the strings by pulling on a tow with monkey fists on the end. Hereby the alignment of the modules is ensured and once in position the structure will be fixed again with a solid pin.

%Beside the in chapter 6 mentioned measures against corroding several sacrificial anodes are additionally attached on the anchor. They are made from zinc and are electrically conductive linked to the strings. The sacrificial anodes protect the important steel parts from damage due to electrochemical reactions by corroding easier than the steel parts and therefore before them. Thus the erosion of the steel in the strings is suppressed until the sacrificial anodes are depleted.}

\section{Qualification and Testing}
\label{sec:testing}
Both module housings of the POCAM and the sDOM were subject to environmental tests to guarantee their use in deep-sea applications. Both housings withstood vibration tests done at IABG\footnote{Industrieanlagen-Betriebsgesellschaft mbH (IABG)}. 
Following a standard test for subsea equipment, the housings with the fully integrated electronics and attached module mounting dummies where exposed to sine vibrations in a frequency sweep from 5 to 150\,Hz at 5\,G in three different axes. Additionally, several shock tests were performed with an acceleration of 10\,G.
%The housings with the fully integrated electronics and attached module mounting dummies where exposed to sine vibrations in a frequency sweep from 5 to 150\,Hz at 5G in three different axes. Additionally several shock tests were performed with an acceleration of 10G.
Furthermore, the new housing of the sDOM was successfully tested in a pressure chamber at the company Nautilus with a pressure sweep up to 385\,bar qualifying for operating at a depth of 2600\,m in the Cascadia Basin.

\begin{figure}[h!]
    \centering
	\includegraphics[height=0.3\textheight]{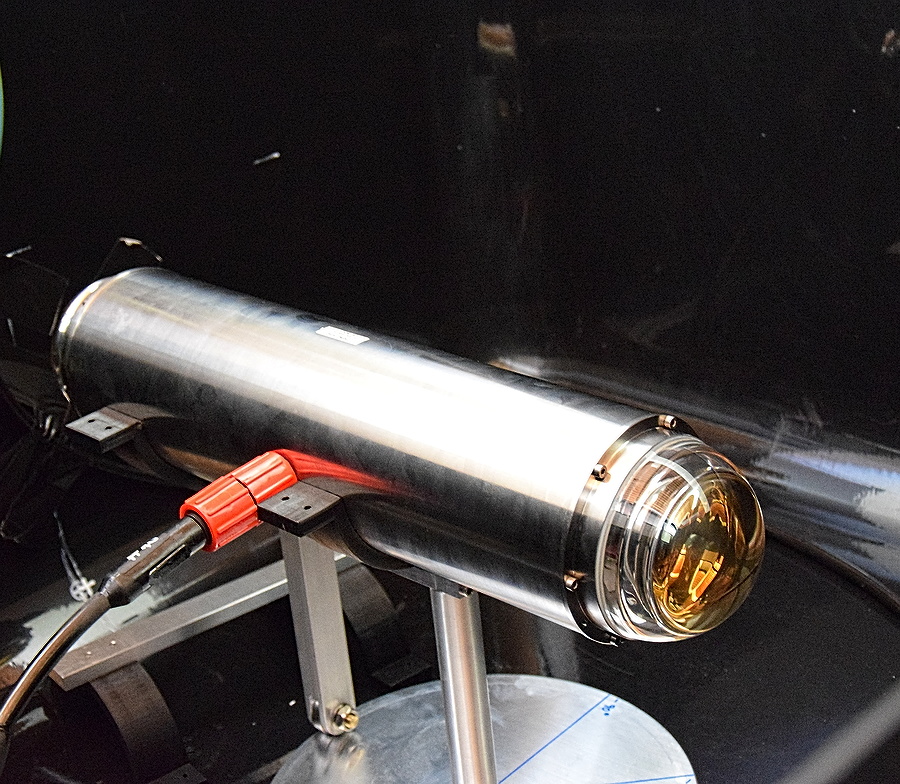}
	\includegraphics[height=0.3\textheight]{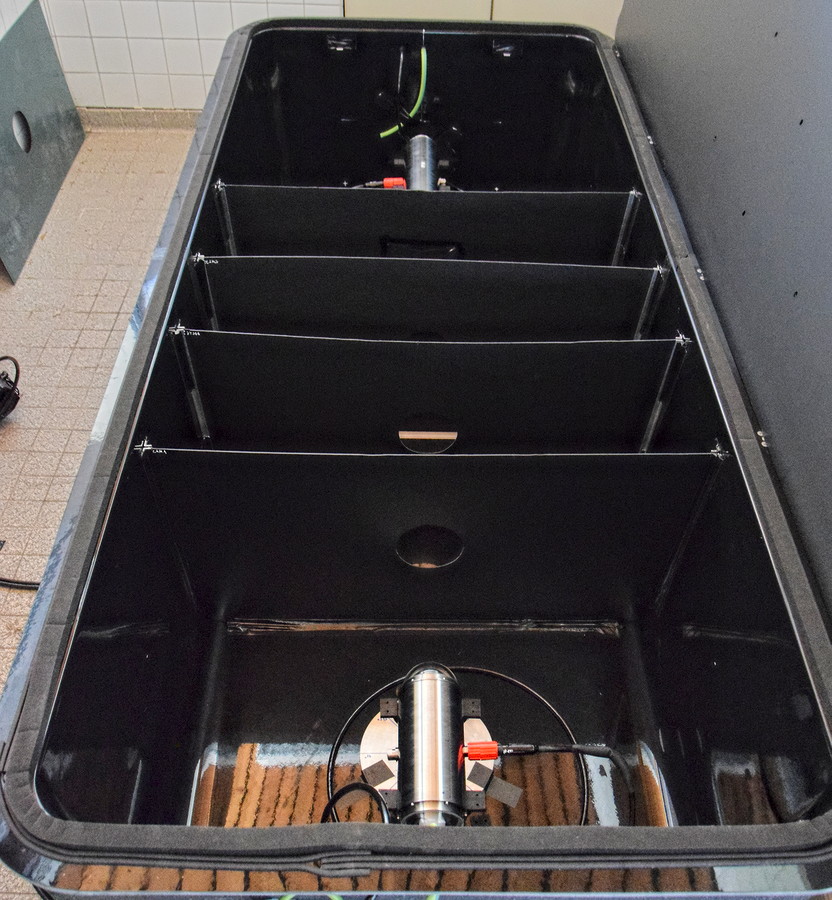}
	\caption{Wet test setup at TUM. The sDOM is fixed to a rotation mount inside a 3\,m $\times$ 1.2\,m pool with baffles for suppressing reflected light. The pool is filled with water during the tests and closed by a light-tight lid.}
	\label{fig:pool_1}
\end{figure}

\begin{figure}[h!]
    \centering
	\includegraphics[width=0.85\textwidth]{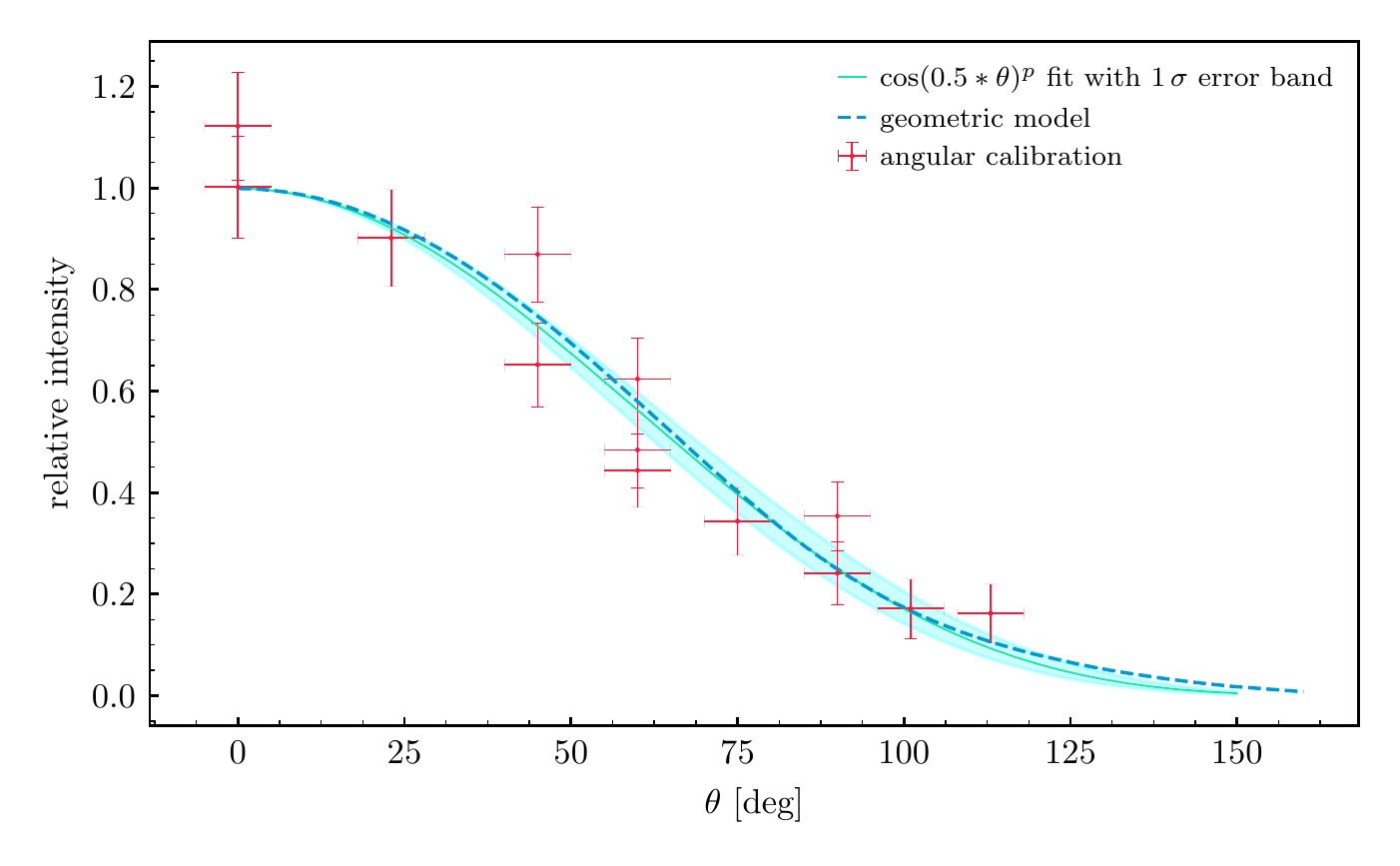}
	\caption{First evaluation of the angular acceptance of the sDOM module submerged inside the water tank. The data have been fitted with a function of the form $\cos(0.5\,\theta)^p$, where $p$ turned out to be $4.0 \pm 0.4$. Also for comparison a pure geometric model of the acceptance is shown, where the projected surface area of the tilted PMT is calculated.}
	\label{fig:pool_scan}
\end{figure}

In order to qualify the angular acceptance of the modules under realistic conditions \rev{they were operated} submerged in water \rev{inside} a custom-built darkened pool, which has been built at TUM\footnote{Technische Universität München} (see Fig.\,\ref{fig:pool_1}). Specific holding structures were designed to place and rotate a POCAM and an sDOM  independently and thus determining the angular acceptance between the two modules. Images of the pool, and the rotation module holder can be seen in Fig.\,\ref{fig:pool_1}. Results from the angular acceptance scan are presented in Fig.\,\ref{fig:pool_scan}. 
%Due to tight time constraints only a rough measurement with large angular uncertainty was possible. This, nevertheless, was enough 
The measurements taken confirm the model expectation based on a geometrical projection of the tilted photocathode area of the PMT. A good analytic description of the angular acceptance is $\cos(0.5\,\theta)^p$, where $p$ has been left free as a fit parameter resulting in a best-fit value of $p = 4.0 \pm 0.4$.
\par
%
%
%\begin{figure}[h!]
%    \centering
%	\includegraphics[angle=270,origin=c,width=0.6\textwidth]{img/pool-complete.JPG}
%	\caption{}
%	\label{fig:pool_2}
%\end{figure}\par
%
%
\par

As part of the testing procedure at the ONC facilities, the modules were tested again individually for any current leakage and ground faults that might have resulted from the shipment to Canada. The modules were operated for about 10 minutes each in a salt-water pool and the ground current was monitored. No issues \rev{were} identified during this test phase. 

Finally, also the fully integrated strings were tested while mounted on the spool in a salt-water test pool at the ONC marine technology center. This test lasted for several days in order to test for water leakage and ensure proper functionality in a salt-water environment.
\rev{While the PMTs have to be protected against overexposure, 3D-printed protective shells covered the glass hemispheres of the sDOMs, and additionally, a tarp canopied the pool.}
%To protect the PMTs against overexposure the glass hemispheres of the sDOMs were covered by 3D-printed protective shells and additionally the pool was covered by an opaque tarp.
The 3D-printed protective shells remained on the modules until right before the deployment. During the pool test the ONC IT team also ran first tests with  driver software that will later be used to operate STRAW and make the measured data available on the ONC website.

\section{Deployment}
\label{sec:deployment}
A top-down approach was chosen for the deployment of STRAW. The mooring lines were transported on custom-made spooling systems. 
\rev{Each spool is powered by an engine, which allows a controlled winding over a wired remote control.}
%Each spool allows to wind one string using an engine controlled by hand. 
%Till the weather conditions are allowing a safe deployment the spooling systems with the strings and the attached modules where stored under deck in order to prevent harm by potentially heavy weather conditions and the PMTs from overexposure by sunlight.
During the deployment and
%the spools were brought into position at the end of the back deck. Here two buoys instead of the designated one were attached to the top spacer out of two reasons: First to prevent the string from lowering to much into the sea during the spooling process and second to attach the heavy lift line for a safe guidance to the seafloor. 
after the connection of the two buoys, a second smaller boat secured the top of the string during the unspooling in water. When the string was almost entirely in water the Mini Junction Boxes and the anchor were mounted at the end of the string. Before final deployment, an electric test of the mounted string was performed.  
Right after the test, the anchor was lifted into the sea via a crane on the back deck. %Now the second boat could drive back to the back deck with the top of the string and the heavy lift line could be connected to the designated buoy while the second buoy got disconnected. 
The whole structure was then lowered in water at a controlled speed of 0.3\,m/s to the sea floor. After the anchor landed on the seabed at a depth of 2600\,m, an acoustic release ensured the disconnection from the heavy lift line. %The same procedure was applied to the second string.\\

\section{First data}
\label{sec:conclusions}
\begin{figure}[h!]
    \centering
	\includegraphics[width=\textwidth]{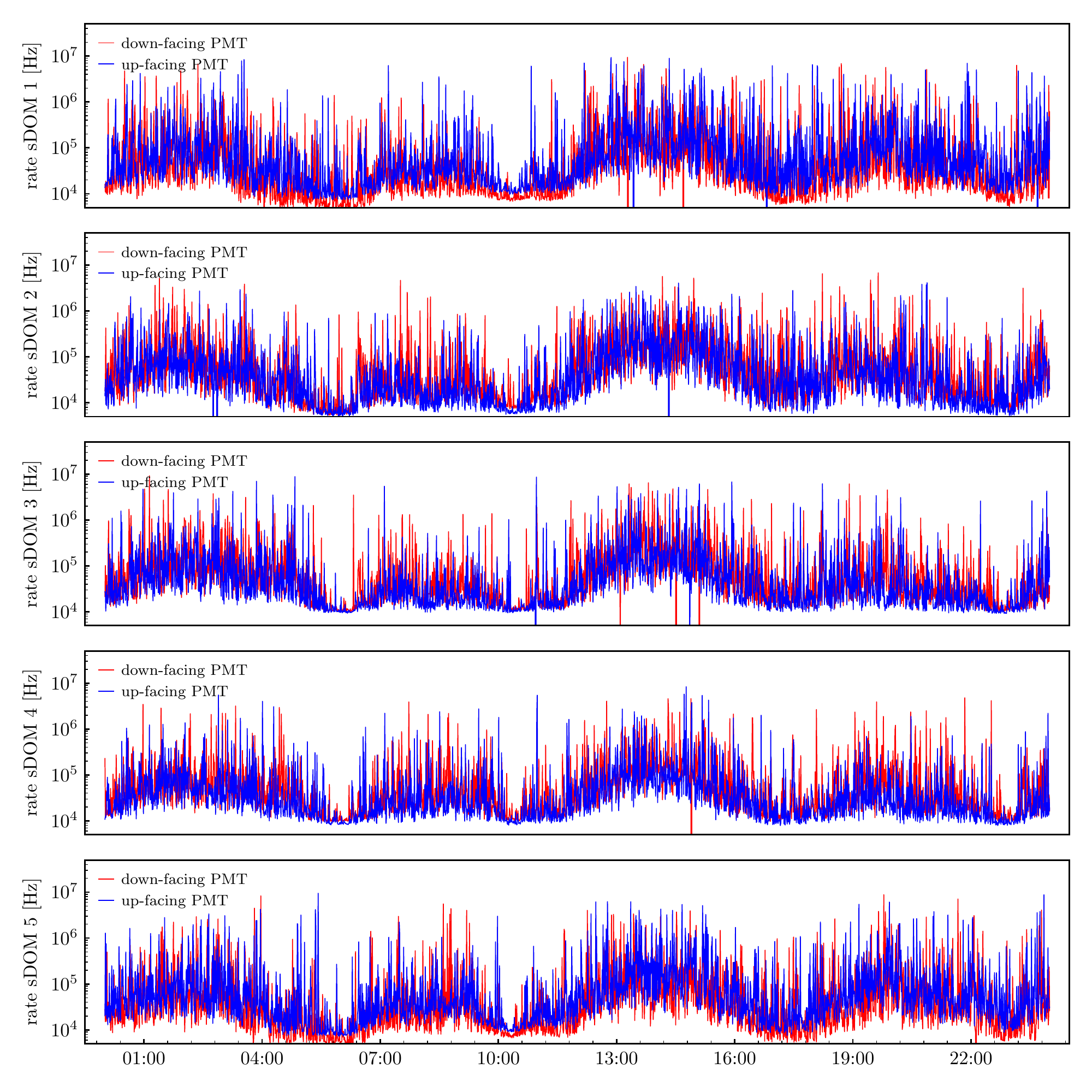}
	\caption{A first glimpse on the dark rate recorded in all five sDOMs throughout one day %(the rates corresponding to the upward facing PMTs are drawn in blue and the ones of the downward facing PMTs in red). 
	The rates are highly variable and range from around 10\,kHz up to several MHz for short periods of times (few seconds). The general trend is the same in all sDOMs with the most plausible explanation for the variable part of the background light being bioluminescence caused by deep-sea animals and microorganisms.}
	\label{fig:biorates}
\end{figure}\par
\begin{figure}[h!]
    \centering
	\includegraphics[width=\textwidth]{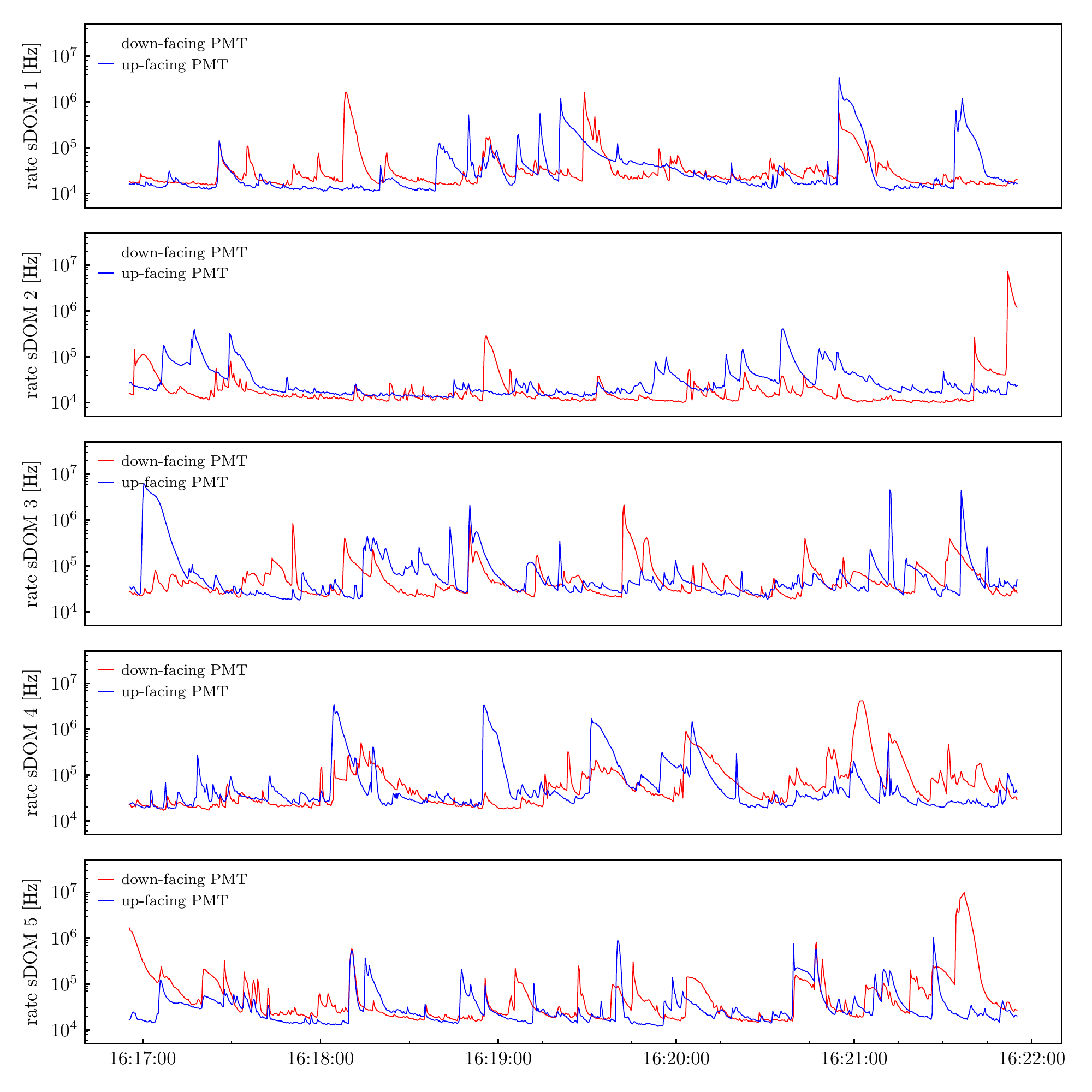}
	\caption{Similar to Fig.\,\ref{fig:biorates}, but rates recorded for only five minutes at a readout rate of 3 Hz. In this representation the fine structure of the flashes most likely caused by luminescence become visible.}
	\label{fig:biorates_fast}
\end{figure}\par
\rev{The final evaluation of the optical site will be subject of a separate publication. In Fig.\,\ref{fig:biorates} we show the rates measured by all the ten deployed sDOM PMTs over the course of 24\,h. The rate varies between ca. 10\,kHz and 100\,kHz with spikes reaching several MHz for a few seconds. The most likely explanation for the variable component of the measured ambient light is the ambient bioluminescence from deep-sea animals and microorganisms \cite{tamburini_deep-sea_2013}. This becomes even more evident when taking a closer look at the finer time structure of the variable background light (Fig.\,\ref{fig:biorates_fast}). This is consistent with previous observations in the Mediterranean Sea and has been proven to be not a limitation for the operation of neutrino telescopes in sea water. A very detailed evaluation of the measurements is currently in preparation.}

\section{Conclusions}
\label{sec:conclusions}
We have reported on the design, the construction and the deployment of the STRAW instrumentation\rev{, as well as presented a very detailed characterization of the emitter and sensor modules and components involved}. STRAW has started the characterization of the deep Pacific Ocean at the Cascadia Basin off the coast of Vancouver Island in Canada. The measurements of the optical properties are a precondition for the possible  construction of a very large volume neutrino telescope where ONC has already wired hundreds of kilometer of the sea floor.
All the sensors deployed are functioning as expected at a depth of 2600\,m b.s.l.. The data-taking  started immediately after the deployment with a phase of commissioning. Since August the STRAW detector takes data in continues mode allowing a commissioning phase in which a set of calibrations have been performed.

\acknowledgments
We thank Stefan Sch\"onert and Claude Vallee for initiating the contact to Ocean Networks Canada and for fruitful discussions. The authors are grateful and appreciative of the support provided by Ocean Networks Canada, an initiative of the University of Victoria funded in part by the Canada Foundation for Innovation. In addition to their whole team, special thanks go to Jeannette Bedard and Ross Timmerman for assisting us in integrating STRAW into the ONC network infrastructure.
This work is supported by the German Research Foundation
through grant SFB\,1258 ``Neutrinos and Dark Matter in Astro- and Particle Physics'' and the cluster of excellence ``Origin and Structure of the Universe''.

\bibliographystyle{plain}
\bibliography{bibliography.bib} 
\end{document}